\documentclass [12pt,article] {article}

\usepackage[T1]{fontenc}
\usepackage[english]{babel}

\usepackage[a4paper,top=1in,bottom=1in,left=1in,right=1in,heightrounded,bindingoffset=0cm]{geometry}

\usepackage{amsmath,amssymb,amsfonts,amsthm,amscd}
\usepackage{bm}

\usepackage{booktabs}
\usepackage{tabularx}
\usepackage{array}

\usepackage{graphicx}
\usepackage{rotating}
\usepackage{epstopdf}

\usepackage{floatrow}
\usepackage{caption}
\usepackage{setspace}
\usepackage{placeins}

\usepackage{natbib}

\usepackage{appendix}
\usepackage{etoolbox}
\usepackage[gen]{eurosym}
\usepackage{lscape}
\usepackage{verbatim}
\usepackage{xcolor}

\usepackage[colorlinks=true,citecolor=blue,linkcolor=blue,urlcolor=blue]{hyperref}

\makeatletter
\def\@fnsymbol#1{\ensuremath{\ifcase#1\or \dagger\or \mathsection\or \ddagger\or
		\or \ddagger\ddagger \else\@ctrerr\fi}}
\makeatother

\title{\textbf{The “Gold Rush” in AI and Robotics Patenting Activity. Do innovation systems have a role?}}

\vspace{0.6cm}

\author{
	Giovanni Guidetti\thanks{Department of Legal Studies, University of Bologna. E-mail: g.guidetti@unibo.it.}
	\and
	Riccardo Leoncini\thanks{Department of Legal Studies, University of Bologna and IRCrES-CNR, Milan. E-mail: riccardo.leoncini@unibo.it.}
	\and
	Mariele Macaluso\thanks{ESOMAS Department, University of Turin. E-mail: mariele.macaluso@unito.it.}
}

\date{}

\begin{document}
\maketitle

\begin{abstract}
\singlespacing
This paper studies patenting trends in artificial intelligence (AI) and robotics from 1980 to 2019. We introduce a novel distinction between traditional robotics and robotics embedding AI functionalities. Using patent data and a time-series econometric approach, we examine whether these domains share common long-run dynamics and how their trajectories differ across major innovation systems.
Three main findings emerge. First, patenting activity in core AI, traditional robots, and AI-enhanced robots follows distinct trajectories, with AI-enhanced robotics accelerating sharply from the early 2010s. Second, structural breaks occur predominantly after 2010, indicating an acceleration in the technological dynamics associated with AI diffusion.
Third, long-run relationships between AI and robotics vary systematically across countries: China exhibits strong integration between core AI and AI-enhanced robots, alongside a substantial contribution from universities and the public sector, whereas the United States displays a more market-oriented patenting structure and weaker integration between AI and robots. Europe, Japan and South Korea show intermediate patterns. 
\vskip0.7cm
\textbf{J.E.L. classification}:  O30, O31, O33
\end{abstract}

\newpage

\section{Introduction}

The rapid expansion of artificial intelligence (AI) and robotics has become a defining feature of contemporary technological competition \citep{UNCTAD2025TIR}. What is increasingly described as an “AI gold rush” reflects not only accelerated inventive activity but also an intensified race for technological leadership, with the US and China at the forefront shaping global innovation trajectories \citep{AIIndex2025}. In the US, \emph{2025 America’s AI Action Plan} emphasises private-sector leadership, deregulation, and capital mobilisation to secure competitive advantage in AI-driven industries. China’s \emph{Global AI Governance Action Plan} (2025), by contrast, embeds AI development within a state-coordinated framework combining industrial policy, strategic planning, and technological sovereignty. Other advanced economies, including the European Union (EU), Japan, and South Korea, position themselves within this landscape through strategies that combine regulatory governance, capability-building, and structured public–private coordination in different ways.
Recent strategies illustrate these institutional differences. The European Union’s \emph{AI Continent Action Plan} (April 9, 2025) combines technological leadership with democratic safeguards and “trustworthy AI,” reflecting a standards-based regulatory model. Japan’s \emph{Basic AI Plan} (December 23, 2025) integrates innovation support with risk management amid concerns about international competitiveness. South Korea’s \emph{National AI Strategy: Policy Directions} (September 26, 2024) advances a coordinated public–private framework to strengthen infrastructure, semiconductor capabilities, and global AI leadership.

The intensity of this technological competition is clearly reflected in patenting dynamics. AI-related patent applications have grown at rates far exceeding the average annual growth of patents across technological fields \citep{WIPO2019}. In particular, between 2013 and 2016, AI applications in robotics and control methods recorded average annual growth rates of approximately 55\% — among the highest observed across patent classes \citep{OECD2019,WIPO2019}. The parallel expansion of robotics-oriented patenting and the rising demand for robotic technologies have been associated with measurable contributions to GDP growth \citep{GraetzMichaels2018,RajSeamans2018}.

A substantial body of economic research examines the effects of AI and robotics on productivity growth, labour demand, task composition, and sectoral exposure \citep[e.g.][]{GraetzMichaels2018,AcemogluRestrepo2020,Kromannetal2020}%
\footnote{This literature highlights broader socio-economic implications, including effects on firm performance, competitiveness, inequality, occupational structure, the organisation of work, and skill demand \citep{BrynjolfssonMcAfee2012,BrynjolfssonMcAfee2014,ArntzEtAl2016,FreyOsborne2017}. A central strand of this research focuses on the impact of AI and robotics on productivity and employment \citep[e.g.,][]{GraetzMichaels2018,Kromannetal2020,Dauthetal2021}, assessing whether these technologies are likely to generate labour market displacement or adjustment \citep[e.g.,][]{AcemogluRestrepo2020,Chiacchioetal2018,GraetzMichaels2018}, and which tasks and skills are most exposed to automation \citep{SquicciariniNachtigall2021,KlenertMaciasAnton2023}.} 
Within this framework, robotics is typically proxied by aggregate measures of automation --most commonly industrial robot adoption -- with robots defined as autonomous programmable machines that perform productive tasks and interact directly with the physical environment \citep{MaciasKlenertAnton2021}. By contrast, AI is analysed either as a distinct domain of inventive activity or as a general-purpose technology characterised by broad complementarities and cross-sectoral diffusion \citep{Trajtenberg2019,MartinelliEtAl2021}. This distinction largely reflects differences in measurement: robotics is observed through adoption data, while AI is identified using patent-based indicators of inventive output.
At the same time, a growing empirical literature has substantially improved the identification of AI- and robotics-related inventions in patent data, through the use of increasingly detailed CPC classifications and text-based retrieval algorithms \citep{BaruffaldiEtAl2020,Samoili2020,LiuEtAl2021,GiczyEtAl2022}. These methodological advances are particularly relevant for AI, whose technological boundaries remain fluid and continuously expanding \citep{Iso2012,OECDb2019,HLEG2019,WIPO2019}.
Yet even with these improvements, AI is generally analysed as a distinct inventive domain, while robotics is treated as a broad automation technology. What remains largely unobserved is the subset of robotic innovation that incorporates AI-based functionalities, such as learning, perception, and adaptive control. Distinguishing between traditional rule-based robotic systems and robotic technologies that embed AI-driven capabilities is therefore crucial, as AI’s economic impact depends not only on its autonomous diffusion but also on its GPT nature encompassing different sectors and technologies in which it becomes embodied and on its complementarities.

Moreover, the evolution of these complementarities is likely to differ across institutional environments. The literature on innovation systems emphasises that technological trajectories are shaped by country-specific configurations of firms, universities, public agencies, and policy frameworks \citep{Lundvall1992,Nelson1993,EtzkowitzLeydesdorff2000}. Innovation is thus understood as an interactive and cumulative process embedded in institutional settings that influence knowledge creation, coordination mechanisms, and capability formation.
This perspective is particularly relevant in the case of AI, because its pervasive nature engenders widespread cross-sector diffusion with porous and evolving boundaries \citep{RighiEtAl2020}. AI develops through interactions across multiple domains and organisational contexts. Accordingly, the trajectory of AI and robotics cannot be understood independently of the environments in which technological change unfolds. The extent to which AI-related knowledge becomes integrated into robotics—and the nature of their co-evolution—may therefore depend on institutional characteristics, including policy priorities and patterns of public–private interaction.
Despite substantial progress in measuring and analysing AI and robotics, an important empirical gap remains. While the economic consequences of these technologies have been widely examined, and patent-based methodologies now permit increasingly precise identification of each domain, the joint evolution of core AI, traditional robots, and robots embedding AI functionalities has not been systematically studied across institutional environments.
This paper addresses this gap by examining two related questions: first, whether traditional robots and robots incorporating AI functionalities can be empirically distinguished as separate technological domains, while preserving a distinct identification of core AI technologies. Second, conditional on this distinction, whether these three domains exhibit common long-run dynamics and how their trajectories differ across national innovation systems.

To address these questions, we construct a patent-family dataset from PATSTAT covering 1980–2019. We combine Cooperative Patent Classification (CPC) codes, keyword-based identification, and document-level text mining to define three mutually exclusive technological domains: core AI, traditional robots, and robots with embedded AI functionalities. Core AI comprises inventions related to foundational AI capabilities—such as learning, reasoning, perception, and autonomous decision-making—rooted in established AI subfields including machine learning and knowledge representation \citep{EPO2017,Samoili2020}. Traditional robots includes industrial, service, and social robots following the taxonomy in \citet{Samoili2020}.
Robots with AI-based technologies (AI-enhanced robots) embed AI structurally and integrally within their capabilities.
This disaggregation yields a measure of inventive activity that isolates AI-driven robotic innovation from conventional automation, enabling a comparative analysis of their technological evolution.
We then examine the long-run dynamics of these series using time-series methods, providing, to our knowledge, the first systematic time-series analysis of the joint evolution of core AI, traditional robots, and AI-enhanced robots. Specifically, we test for non-stationarity, identify structural breaks, and estimate pairwise cointegration relationships across technological domains and countries. This empirical framework allows us to assess whether core AI, traditional robots, and AI-enhanced robots share common stochastic trends or follow distinct technological trajectories, and whether these patterns differ across institutional environments.
By separating core AI from its application in robotics and analysing their long-run relationships across national innovation systems, the paper provides a systematic account of how AI-driven complementarities emerge and evolve within applied technologies.

Our findings highlight three main results. First, patenting activity in the three domains exhibits distinct long-run dynamics. Core AI expands rapidly, though unevenly across technological classes; traditional robots follows a more gradual trajectory; and AI-enhanced robots displays a marked acceleration beginning in the early 2010s.
Second, structural breaks are concentrated in the post-2010 period, consistent with a shift in innovation regimes associated with the diffusion of AI-related capabilities into robots. Third, long-run relationships vary systematically across countries and depend critically on distinguishing AI-enhanced from traditional robots. China exhibits strong cointegration between core AI and AI-enhanced robots, whereas the United States shows limited integration at the applicant level. Europe and Japan display intermediate and technology-specific patterns.

The paper is structured as follows. Section~\ref{sec:literature} discusses the theoretical background of the paper and the main hypothesis to be empirically tested. Section~\ref{sec:data} describes the data, reports some descriptive analysis, and presents the empirical strategy. Section~\ref{sec:results} reports and discusses the results. Section~\ref{sec:conclusion} provides some concluding remarks.

\section{Conceptual Framework}\label{sec:literature}

Innovation is strongly associated with economic competitiveness \citep[e.g.][]{Castellacci2008,ClarkGuy1998}, and patents are widely used as a proxy for innovative output \citep{Mansfield1989,HallJaffeTrajtenberg2001,BessenMeurer2008,Acemogluetal2011,Montobbioetal2022,EckertLanginier2014,Cominoelal2019}. A large empirical literature has relied on patent data to study the dynamics of technological change and its economic effects across countries, sectors, and time. While alternative appropriability mechanisms, such as trade secrets or the strategic use of patenting, play an important role in firms’ innovation strategies \citep{CohenNelsonWalsh2000,Hall2004,CohenGurunKominers2016}, patents remain a particularly informative source for analysing emerging and fast-evolving technologies.

This is especially true in the case of artificial intelligence (AI) for two main reasons. First, AI-driven innovation is expected to generate wide-ranging transformative effects on economies and societies, unfolding amid high uncertainty, rapid experimentation, and intense competition over standards and dominant designs. In such contexts, patenting activity provides a timely signal of inventive efforts and technological positioning. Second, AI technologies are highly pervasive, spanning multiple sectors and applications, making patent protection central to appropriability and the control of knowledge diffusion across technological fields. These characteristics distinguish AI from more mature technologies and reinforce the relevance of patent data for identifying technological trajectories.

For similar reasons, patent data are also particularly informative for the study of robotics, as they allow the analysis of inventive activity at its source and capture the technological origins of robotic innovation independently of its downstream adoption. Since the 1980s, empirical research on robotics has predominantly focused on ``traditional'' robots, mirroring the rapid growth of robotics-related innovative activity over that period \citep[e.g.][]{TsengTing2013}. Much of this literature relies on aggregate measures of robot adoption, which have been extensively used to study the economic and employment effects of traditional robots, particularly in manufacturing industries such as automotive, electronics, and chemicals.%
\footnote{For instance, the International Federation of Robotics produces widely used aggregate data for robot adoption at the country level \citep[e.g.][]{Chiacchioetal2018,GraetzMichaels2018,AcemogluRestrepo2020b}.}
While this approach has generated important insights into the demand-side impacts of robotics, it has left relatively unexplored the sectors in which robotic innovations originate and the technological processes underlying their development \citep{Montobbioetal2022}. 

At the same time, the increasing pervasiveness of artificial intelligence has progressively blurred the boundaries between robotics and other digital technologies, contributing to the emergence of a new generation of robots that incorporate AI capabilities \citep{OECD2019,WIPO2019,Samoili2020}. In this context, aggregate indicators of robot adoption are informative about diffusion patterns and downstream economic effects. Yet, they do not allow a systematic distinction between conventional robotic systems and AI-enhanced robots. This limitation is particularly relevant given the heterogeneity of robotic technologies and the importance of their technological origins along the innovation process, both at the patent level and at the country (cluster) level.

The pervasivity of AI has suggested that it could be framed as a GPT \citep{Trajtenberg2019}, with the development of related technologies, including robotics, embedded in the broader transformation associated with Industry~4.0 \citep{MartinelliEtAl2021,BenassiGrinzaRentocchini2020}. At the same time, more recent contributions debate whether AI should be interpreted as the emergence of a new technological paradigm or as a radical transformation within the existing ICT trajectory \citep{DamioliEtAl2025}. Taken together, these perspectives underscore the pervasive character of AI-driven technological change.

However, such a change cannot be fully understood without explicitly considering the institutional and policy environments in which AI and robotics co-evolve. Indeed, AI-related innovation unfolds across multiple technological domains and organisational settings \citep{RighiEtAl2020,LiuEtAl2021,MarianiEtAl2023}, and their development and implications for robotisation depend crucially on interactions between private and public actors within policy frameworks that steer techno-economic change in strategic directions \citep{NarayananEtAl2022}. 

A hint at the different roles of the institutional framework is presented in Table \ref{inst_fram}, in which the different roles of the state and/or of the private firms in generating AI patenting activity emerges quite clearly by confronting, for example, China and the US.
In China, the government pulls firms and institutions and, in general, controls the innovative processes and the industrial dynamics by favouring and fostering partnerships between tech giants and incumbents within industrial sectors to promote firms’ digitalisation. In contrast, in the US, although the government implements strategies to push and boost firms and institutions, the structural dynamics are mainly governed by market mechanisms in which tech giants and new tech players push away incumbents who do not manage to adjust to new technologies. 

Within this perspective, innovation is understood as the outcome of interactive and cumulative learning processes embedded in broader institutional and organisational contexts, rather than as the result of isolated firm-level decisions \citep[e.g.][]{MusiolikEtAl2020}. Knowledge is generated through processes of interactive learning and diffuses through interactions among firms (suppliers, customers, competitors), public and private universities, research institutes, non-profit and state organisations \citep{EtzkowitzLeydesdorff2000}.
Cross-country heterogeneity in technological development is thus the outcome of institutional environments and coordinated learning processes \citep{OECD2002,CastellacciNatera2015}. Differences in innovation performance across countries reflect heterogeneity in institutional configurations shaping knowledge creation, coordination mechanisms, and technological learning. These differences have given rise to distinct national trajectories, as illustrated by the U.S. system \citep{Mowery1992,Mowery1998}, the European experience \citep{Borras2004,CarayannisKorres2013}, South Korea \citep{LeePark2006}, and China \citep{LiuWhite2001,ZhouLeydesdorff2006}.
A key implication of this perspective is that countries differ systematically in their innovation performance because they differ in production structures, knowledge bases, and institutional infrastructures.
Institutional complementarities and flexibility can further support qualitative shifts in innovation dynamics and the emergence of new technological systems \citep{LeonciniMontresor2003}, thereby strengthening the development of technological capabilities. 

Within this framework, AI, traditional robots, and AI-enhanced robots can be interpreted as interrelated technological domains whose evolution depends on institutional complementarities rather than as independent innovation processes. This perspective motivates our empirical analysis of the long-run evolution of patenting activity in AI, traditional robots, and AI-enhanced robots across major national innovation systems. 

Against this background, our contribution aims to fill the following gaps in the literature. On the one hand, as already pointed out, the literature on the relationships between AI and robots is mainly based on aggregate data on robot production and diffusion, we realise that to assess the way in which AI as a GPT influences the robots technology, first of all, we need to understand which robots are capable of working under AI supervision.
Hence, on the one hand, our contribution is about the introduction of a novel patent-based framework that explicitly differentiates between AI, traditional robots, and AI-enhanced robots. By combining detailed patent classification codes with keyword-based and text-mining techniques, we construct a dataset that allows us to track these technologies at the patent level separately and to analyse their evolution patterns across countries and over time. This approach enables us to move beyond aggregate measures of robot adoption and to directly examine how AI and robots co-evolve, whether their innovation trajectories converge or diverge, and how these patterns vary across national contexts. 
On the other hand, we are interested in the long-run dynamics of both AI and robots, and in this regard, we reckon that the use of time series analysis might be the proper tool to analyse the separate long-run dynamics of AI and robots, as well as the interactions among the
As for the former, we use stationarity analysis to understand how the series has evolved over time and complement it with tests for structural breaks.
In this way, we can characterise how differently AI and robots have evolved over the long run since their very start.
The analysis of structural breaks will allow us to assess when and how some institutional factors were contemporaneous with the break, and will give us an idea of the eventual presence of exogenous factors that may have influenced or caused the break.
As for the latter, we perform cointegration analysis to understand whether and how AI and robots co-evolved in the long run.
The analysis of cointegration will give us a picture of how, across countries, AI and robots co-moved, highlighting the possible interrelationships that shaped different evolutionary paths within and across countries and patent classes.
These different patterns could be interpreted as the results of interactions among the institutional actors within each system.
More broadly, our framework provides a foundation for studying the interaction between AI and robotics as a system-level phenomenon. It allows us to interpret observed differences in patenting dynamics not only as the outcome of firm-level innovation incentives, but also as the reflection of broader institutional and policy environments shaping technological change.

\section{Data Source and Empirical Strategy}\label{sec:data}

Our analysis relies on the EPO Worldwide Patent Statistical Database (PATSTAT, Spring 2021 edition), one of the most comprehensive and widely used sources for studying technological innovation \citep{DeRassenfosseEtAl2014}. PATSTAT collects harmonised data from over 90 national and regional patent offices, providing detailed, internationally comparable, longitudinal information on patent applications \citep{KangTarasconi2016}. Its broad coverage and standardised structure make it a benchmark dataset for empirical research on emerging technologies, including artificial intelligence and robotics.

The database provides rich bibliographic and technical information, including Cooperative Patent Classification (CPC) codes\footnote{The CPC system categorizes patents based on their technological content and is organized hierarchically: sections (1 letter) represent broad technological areas (i.e., A – Human Necessities; B – Performing Operations; Transporting; C – Chemistry; Metallurgy; D – Textiles; Paper; E – Fixed Constructions; F – Mechanical Engineering; Lighting; Heating; Weapons; Blasting; G – Physics; H – Electricity; and Y – General tagging of emerging cross-sectional technologies, which includes new technological fields and multidisciplinary inventions), classes (1 letter + 2 digits) subdivide sections into specific fields, subclasses (1 letter + 2 digits + 1 letter) further refine the technology, groups (1 letter + 2 digits + 1 letter + 1–3 digits) indicate particular technical aspects within a subclass, and subgroups (up to 11 characters in total) provide the finest level of detail, enabling precise indexing and systematic patent analysis.}, titles and abstracts, application numbers, filing and publication dates, grant status, and citation data (forward and backward). It also contains information on applicants and inventors, such as country of origin and sector of activity, as well as the technological domain in which patents are classified.
By combining these variables, our dataset enables a systematic analysis of AI and robotics patent activity related technologies. Moreover, it allows us to distinguish between AI-enhanced and traditional robots.

\subsection{Data Collection}

Our first contribution is the construction of a novel dataset that distinguishes between AI patents, traditional robot patents, and AI-enhanced robot patents. This data structure allows us to separately identify AI as a GPT and its specific integration into robotic systems, thereby extending existing approaches that either focus on robotics as a homogeneous field or conflate AI-driven innovations with broader automation technologies. Achieving this distinction empirically requires a careful account of technological boundaries, which, in turn, calls for a data construction strategy that ensures both high recall and high precision in selecting relevant patents. To this end, we adopt a combined approach that integrates technology classification codes with targeted keyword-based searches, allowing us to systematically identify inventions related to AI while limiting noise from unrelated technological domains.

A key methodological challenge in the empirical analysis of emerging technologies concerns the reliable identification of relevant patents. In the case of AI, this task is complicated by the absence of a universally accepted definition of AI patenting, as well as by the field’s rapid expansion, fluid technological boundaries, and inherently multidisciplinary nature \citep{LiuEtAl2021,GiczyEtAl2022}.

Building on the notion of core artificial intelligence, AI patents are defined as inventions related to fundamental AI capabilities — such as learning, reasoning, perception, and autonomous decision-making — rooted in core AI scientific subdomains, including knowledge representation,  reasoning and machine learning \citep{EPO2017, Samoili2020}. To operationalise the construction of the AI patent dataset, we first rely on CPC codes retrieved from multiple authoritative sources, including the \textit{Intellectual Property Office} \citep{IPO2019}, and the \textit{World Intellectual Property Organization} \citep{WIPO2019}. The CPC lists provided by \textit{WIPO} and \textit{IPO} are based on a combined use of classification codes and targeted keyword queries, allowing for a broad coverage of AI-related patents while limiting the inclusion of unrelated technological domains (see Table A1 in the Appendix). The complete set of CPC codes employed in the analysis is reported in Table A2 in the Appendix.

We then distinguish between traditional robots and AI-enhanced robots. Existing studies have generally classified robotic technologies either by function -- such as industrial, service, or social robots -- or by their distribution across countries and sectors of application \citep[e.g.][]{MaciasKlenertAnton2021,MartinelliEtAl2021}. Although informative, these classifications abstract from the technological change introduced by the integration of artificial intelligence into robotic systems.
Our approach instead differentiates between robotic inventions that rely on pre-programmed, rule-based operations and those that explicitly integrate artificial intelligence to perform cognitive, adaptive, or autonomous functions. This distinction is relevant because the technological, economic, and societal implications of AI-enhanced robots are likely to differ substantially from those of traditional automation. This questions the standard practice in the literature of treating robotics as a homogeneous technology \citep[e.g.][]{GraetzMichaels2018,AcemogluRestrepo2020b,OECD2019,WIPO2019}.

To implement this distinction, we adopted a two-step approach. 
First, we complement CPC-based searches with keyword searches in patent titles and abstracts. For this purpose, we rely on the robotics-related keyword list developed by \citet{Samoili2020}, which covers intrinsic AI subdivisions in robotic process automation. In this framework, the integration of AI into robotics encompasses the design and use of intelligent technological systems that support or replace human tasks, extend feasible actions beyond human capabilities, and improve efficiency by alleviating technical constraints and reducing labour or production costs. Within this framework, we identify robot-related patents by retrieving ‘industrial robot’, ‘service robot’, and ‘social robot’ and merging them with patents obtained under the broader ‘robot system’ query. We then identify AI-enhanced robots by augmenting the search with a restricted set of AI-intrinsic keywords—i.e., terms identifying active AI agents—that directly relate to artificial intelligence in robotic systems, including ‘cognitive system’ and ‘control theory’. We retain only matches that occur within the robot patent universe, excluding cases in which these AI-intrinsic terms appear outside robotics. The keyword selection is based on their co-occurrence with core robot terms and their close conceptual link to AI functionality in robotics. This approach prioritises precision by limiting the inclusion of patents unlikely to involve AI-enhanced robotic systems, thereby reducing false positives. Accordingly, we restrict the baseline definition to AI-intrinsic terms to limit classification noise.

Second, to further refine the identification of AI-enhanced robots and capture AI-related content that may not be fully identified by the initial keyword filtering, we apply a text-mining procedure using a Python keyword-matching algorithm. The text-screening procedure is applied to robot-related patent families identified in the initial extraction step, covering `industrial robots' (25,326), `service robots' (6,179), `social robots' (117), `robot systems' (59,268). Patent titles and abstracts are imported and split into individual sentences, which are then scanned for a predefined set of AI-related keywords in a case-insensitive manner. The keyword list is compiled from multiple authoritative sources \citep{IPO2019,BaruffaldiEtAl2020,Samoili2020}.
For each patent, the algorithm records the detected keywords and appends them to a new column in the dataset. We additionally extract word frequencies for both individual keywords and relevant keyword combinations. 

\begin{figure}[h!]
	\centering
	\caption{Distribution of AI-related Keywords in Patent Titles, by Technological Domain}
	\label{fig:keywordstitles}
	\includegraphics[width=0.85\linewidth]{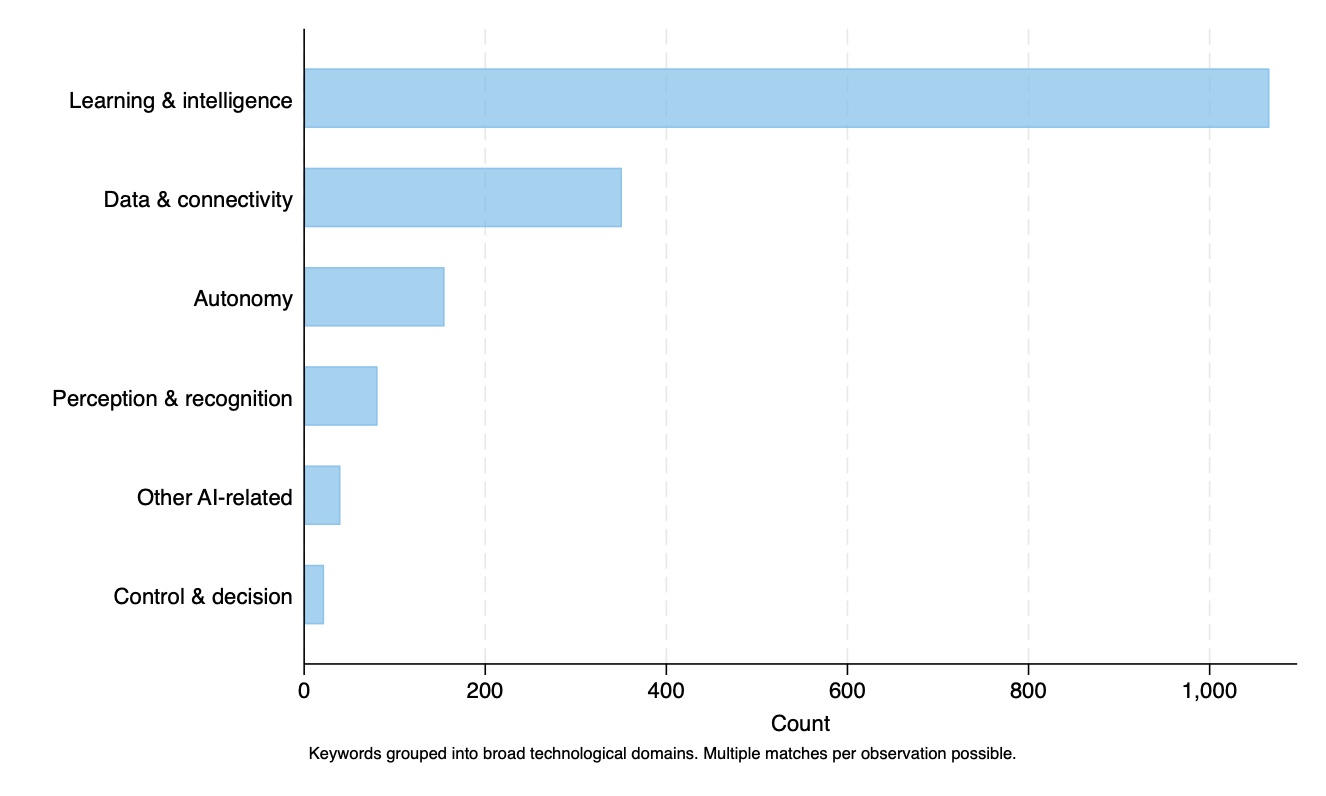}
	\floatfoot{\scriptsize
		\begin{minipage}{0.8\linewidth}
			\emph{Notes:} Keywords extracted from patent titles are grouped into broad technological domains. 
			\emph{Learning \& Intelligence} includes learning, machine learning, neural network, intelligent, intelligence, cognitive, and reasoning; 
			\emph{Perception \& Recognition} includes computer vision, image processing, image recognition, face recognition, speech recognition, and gesture recognition; 
			\emph{Data \& Connectivity} includes cloud, big data, and Internet of Things; 
			\emph{Control \& Decision} includes decision support, prediction, fuzzy logic, and fuzzy control; 
			\emph{Robotics \& Autonomy} includes autonomous robot, autonomous mobile robot, robot system, swarm robot, and unmanned aerial vehicle (UAV); 
			\emph{Other AI-related} includes augmented reality, virtual reality, sentiment analysis, game theory, and genetic algorithm. 
			Keyword categories are not mutually exclusive.
		\end{minipage}
	}
\end{figure}

\begin{figure}[h!]
	\centering
	\caption{Distribution of AI-related Keywords in Patent Abstracts, by Technological Domain}
	\label{fig:keywordsabstract}
	\includegraphics[width=0.85\linewidth]{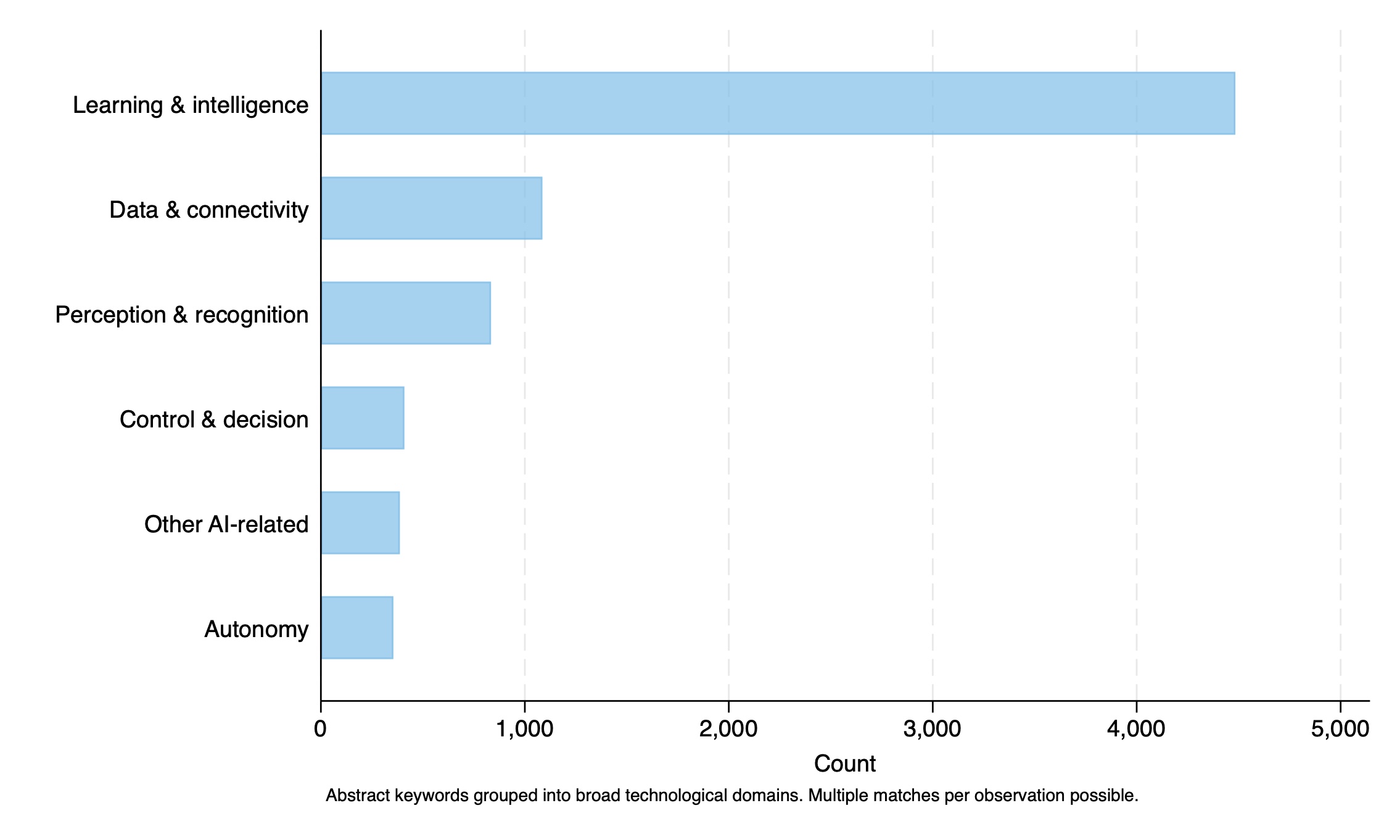}
	\floatfoot{\scriptsize
		\begin{minipage}{0.8\linewidth}
			\emph{Notes:} Keywords extracted from patent abstracts are grouped into broad technological domains using keyword matching. 
			\emph{Learning \& Intelligence} includes learning, machine learning, neural network, deep learning, cognitive, intelligence, and reasoning. 
			\emph{Perception \& Recognition} includes computer vision, image processing, image recognition, face recognition, speech processing, sensor, and LiDAR. 
			\emph{Data \& Connectivity} includes cloud, big data, data mining, Internet of Things, and network data. 
			\emph{Control \& Decision} includes decision support, prediction, control theory, and fuzzy logic. 
			\emph{Robotics \& Autonomy} includes robot, robotic, autonomous robot, autonomous driving, unmanned aerial vehicle, and swarm robot. 
			\emph{Other AI-related} includes augmented reality, virtual reality, genetic algorithm, and anomaly detection. 
			Multiple domains may be assigned to the same abstract.
		\end{minipage}
	}
\end{figure}

To illustrate the technological content captured by this procedure within industrial, service, social robots and robot systems, Figure \ref{fig:keywordstitles} and Figure \ref{fig:keywordsabstract} report the distribution of keywords related to AI extracted from patent titles and abstracts, respectively, grouped into broad technological domains. In both cases, learning and intelligence-related concepts account for the largest share of keyword occurrences, indicating that AI-enhanced functionalities in robots are predominantly associated with learning and inference capabilities. Data and connectivity technologies represent a second important component, while perception and recognition-related terms appear less frequently. Keywords explicitly related to autonomy and control account for a smaller fraction, suggesting that AI capabilities in robots are often embedded in learning and data-processing components rather than framed in terms of stand-alone control architectures.
Based on the information obtained from the text-mining procedure and the co-occurrence-based selection of AI-intrinsic keywords described above, we construct the variable `intelligent', which takes the value 1 if a patent contains at least one AI-related keyword.

This classification approach is explicitly designed to maximise recall at early stages. Broad CPC- and keyword-based searches, combined with sentence-level text mining using comprehensive AI-related dictionaries, ensure an inclusive retrieval of potentially relevant patents. More restrictive, AI-intrinsic keywords are incorporated directly into the extraction process to prioritise precision in the baseline definition.
This procedure, summarised in Table A3 in the Appendix, allows us to distinguish between traditional robots and AI-enhanced robots clearly. By doing so, it addresses key definitional challenges and enables an empirical analysis of whether the diffusion of AI follows patterns similar to, or distinct from, those of conventional robots.

\subsection{Data Construction and Final Dataset} 

Patent data are organised by application year, which provides a closer approximation of the period of invention development than the grant year \citep{LiuEtAl2021}. Following the literature \citep{HallHelmers2013,BarbieriMarzucchiRizzo2020}, we use the patent family—defined as the set of patents filed in different jurisdictions to protect the same invention—as the unit of analysis to avoid double-counting when an invention is filed in multiple countries. To identify the patent family, we primarily consider the maximum forward citation within the family—that is, the number of times a patent is cited by subsequent patents as prior art; if this information is missing, we use the maximum backward citation, which refers to the number of earlier patents cited by the patent itself \citep[e.g.][]{Verhoevenetal2016,BarbieriMarzucchiRizzo2020}.

Our analysis distinguishes between the authority country, where the patent is filed, and the applicant’s country, defined as the country where the entity or individual that filed the patent is legally registered or resides. Authority countries were coded and grouped into six regions: the US, China, Japan, South Korea, Europe, and the rest of the world. Each patent family is attributed to the country of the most relevant patent. To ensure robustness, patent counts are then aggregated at the authority–year level and normalised by total patenting activity in the corresponding year, yielding relative weights that account for variations in aggregate patent volumes over time.

Regarding applicants, PATSTAT contains significant gaps in country information, particularly for China, South Korea, and Japan. While standard procedures, such as those used by the EPO, can probabilistically assign missing countries—for example, patents filed in China with no corresponding families elsewhere are almost certainly (99.9\%) associated with Chinese applicants—in our study, we manually verified and completed all missing applicant countries by searching individually in EPO Espacenet. This careful verification ensures accurate and consistent applicant information, although the sample size is slightly reduced because it relies exclusively on confirmed entries from PATSTAT.

The final dataset consists of 257,680 AI-related patent families, 50,733 traditional robot patent families, and 7,626 AI-enhanced robot patent families. It contains CPC classifications (from broad classes to fine-grained subgroups), patent metadata (ID, family, title, abstract, filing/priority/publication year, kind, IPR, authority, claims, grants), and applicant/inventor information (name, sector, country). 

Patent data are obtained for the period 1980–2019 to identify patents related to AI and robotics. To address potential truncation bias induced by publication and administrative processing delays, which may result in systematically under-recorded patent counts in the most recent filing year, the empirical analysis excludes 2019 and focuses on the period 1980–2018.

The dataset is structured as a time series, organised by classification, authority, and applicant country, enabling consistent, comparable analysis across AI and robot patents.  In our time-series analysis, we focus on patents classified in the following sections: A (Human necessities), B (Performing operations; Transporting), E (Construction), F (Mechanical engineering; Lighting; Heating; Weapons; Blasting), G (Physics), H (Electricity) and Y (General tagging of new technological developments). These sections cover the broad technological spectrum where AI- and robots-related innovations are most likely to emerge, ensuring that our dataset captures both core computational advances and their applications across diverse industrial and societal domains.

\subsection{Descriptive Statistics}\label{para:aioverview}

\begin{figure}[h!]
	\centering
	\caption{Total Number of Patent Families by Filing Year}
	\label{fig:patentfamiliesbyyear}
	\includegraphics[width=0.85\linewidth]{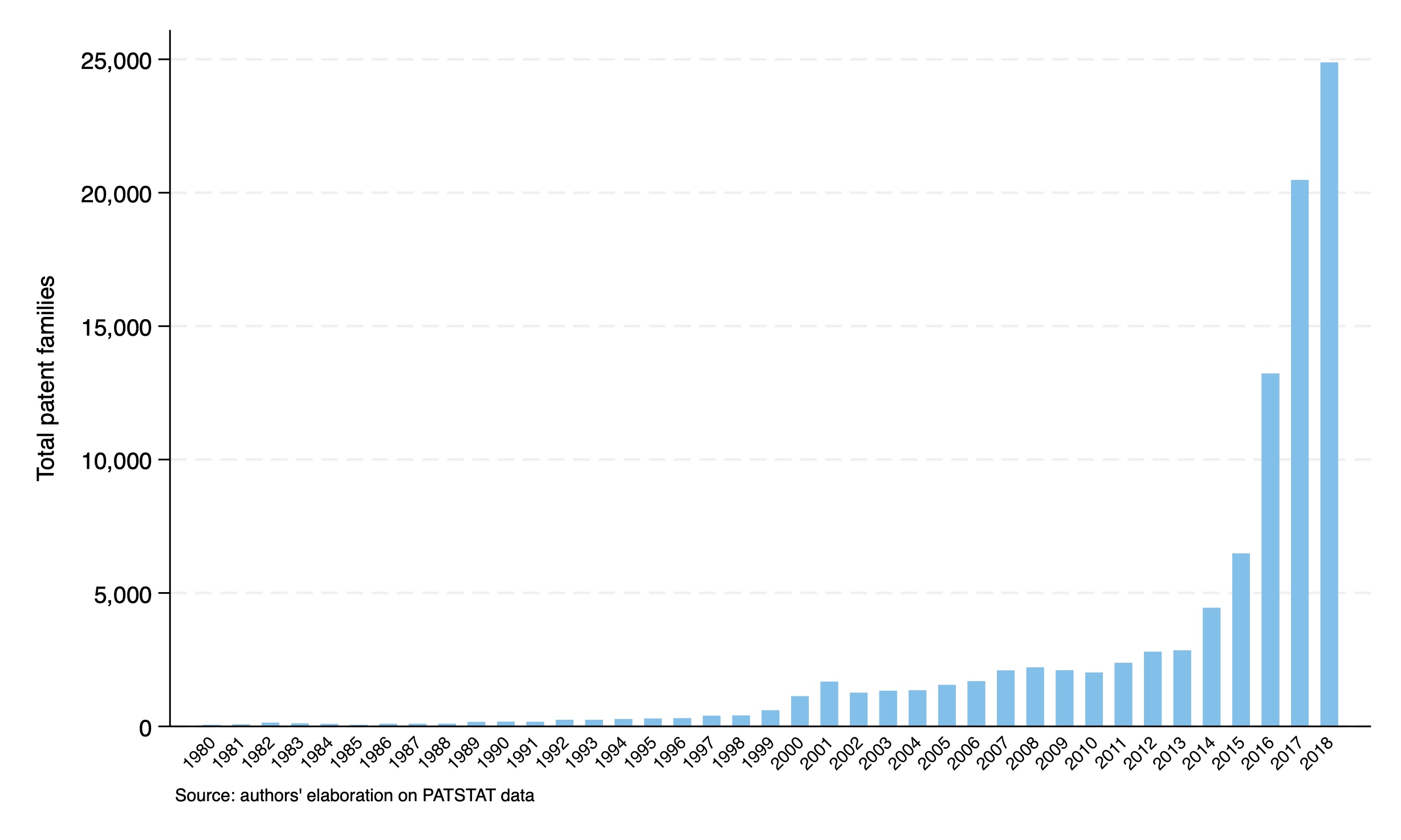}
		\floatfoot{\scriptsize
		\begin{minipage}{0.75\linewidth}
			\emph{Notes:} The figure reports the annual number of AI patent families aggregated by filing year. 
			Patent families are defined following the PATSTAT family identifier and are used as the unit of analysis to proxy distinct inventive outputs, thereby avoiding multiple counting of the same invention across jurisdictions. 
			Observations for 2019 are excluded due to incomplete reporting and truncation in the most recent filing year.
		\end{minipage}
	}
	
\end{figure}

Figure~\ref{fig:patentfamiliesbyyear} shows that patenting activity related to artificial intelligence has grown rapidly over time: more than 25,000 patent applications for inventions related to AI were filed in 2018, and the evolution of AI patent families displays particularly strong increases in the most recent years. At the same time, the long-run pattern of AI patenting reflects a highly uneven distribution across technological domains. 

\begin{figure}[h!]
	\centering
	\caption{Evolution of Patent Stocks by CPC Section}
	\label{fig:stockailogsections}
	\includegraphics[width=0.85\linewidth]{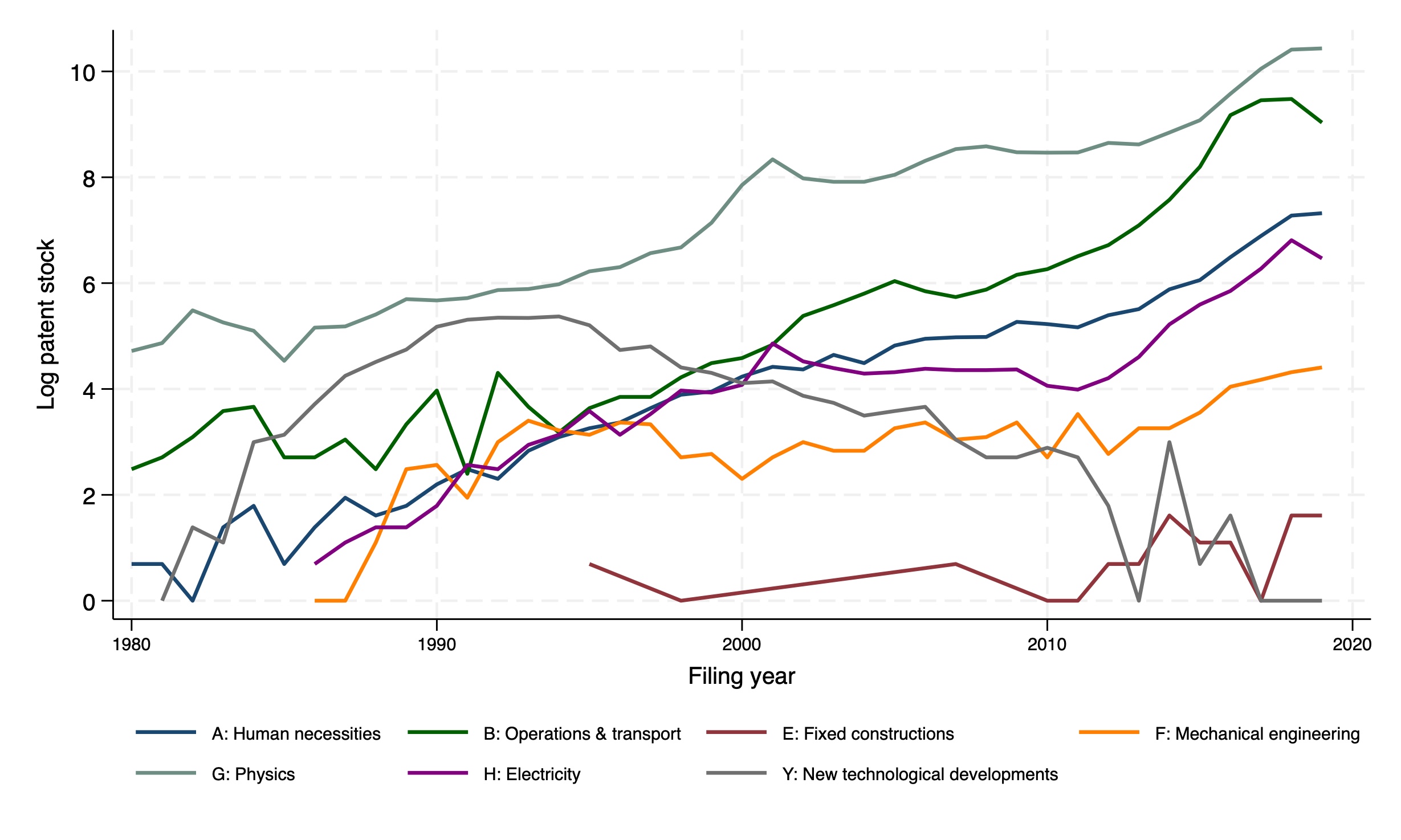}
	\floatfoot{\scriptsize
		\begin{minipage}{0.75\linewidth}
			\emph{Notes:} The figure plots the logarithm of AI-related patent stock by CPC section and filing year (1980--2019). 
			Patent stock is measured using patent families and aggregated at the section--year level. 
			Each line represents the evolution of the log of the patent stock within a given CPC section.
		\end{minipage}
	}
\end{figure}

This heterogeneity is illustrated in Figure~\ref{fig:stockailogsections}, which reports the evolution of the log stock of AI patent families across CPC sections. While patenting activity increases across all sections over time, growth trajectories differ markedly across technological domains: sections related to electricity and physics exhibit earlier and steeper increases, whereas other sections display more gradual growth. Overall, the figure highlights the uneven expansion of AI patenting across technological fields.

\begin{figure}[h!]
	\centering
	\caption{Robot-Related Patent Families by Filing Year}
	\label{fig:robotpatentfamilybyyear}
	\includegraphics[width=0.85\linewidth]{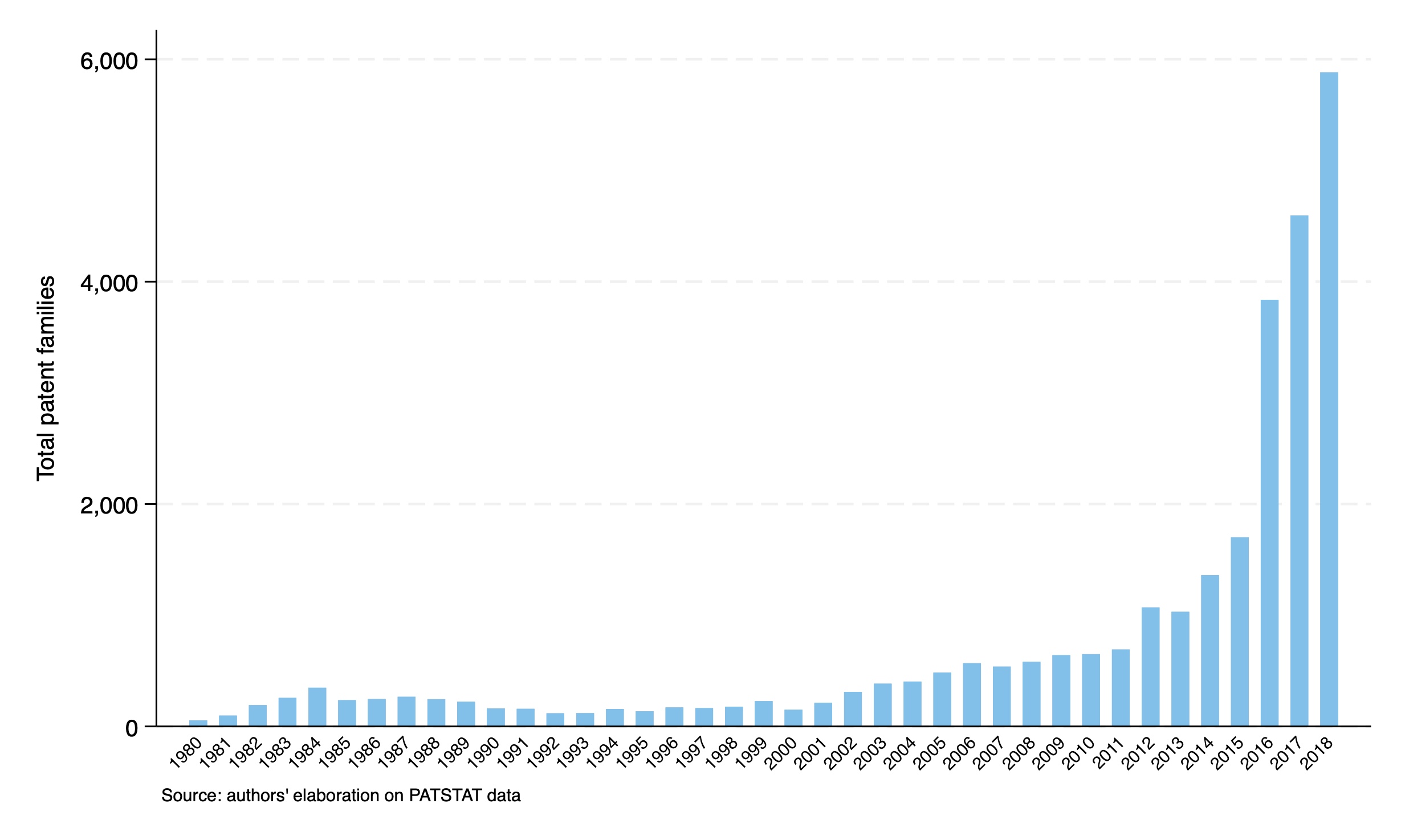}
	\floatfoot{\scriptsize
		\begin{minipage}{0.75\linewidth}
			\emph{Notes:} The figure reports the annual number of total robot patent families aggregated by filing year. 
			Patent families are defined following the PATSTAT family identifier and are used as the unit of analysis to proxy distinct inventive outputs, thereby avoiding multiple counting of the same invention across jurisdictions. 
			Observations for 2019 are excluded due to incomplete reporting and truncation in the most recent filing year.
		\end{minipage}
	}
	
\end{figure}

\begin{figure}[h!]
	\centering
	\caption{Evolution of Patent Stocks: Traditional vs AI-Enhanced Robots}
	\label{fig:stockrobotslog}
	\includegraphics[width=0.7\linewidth]{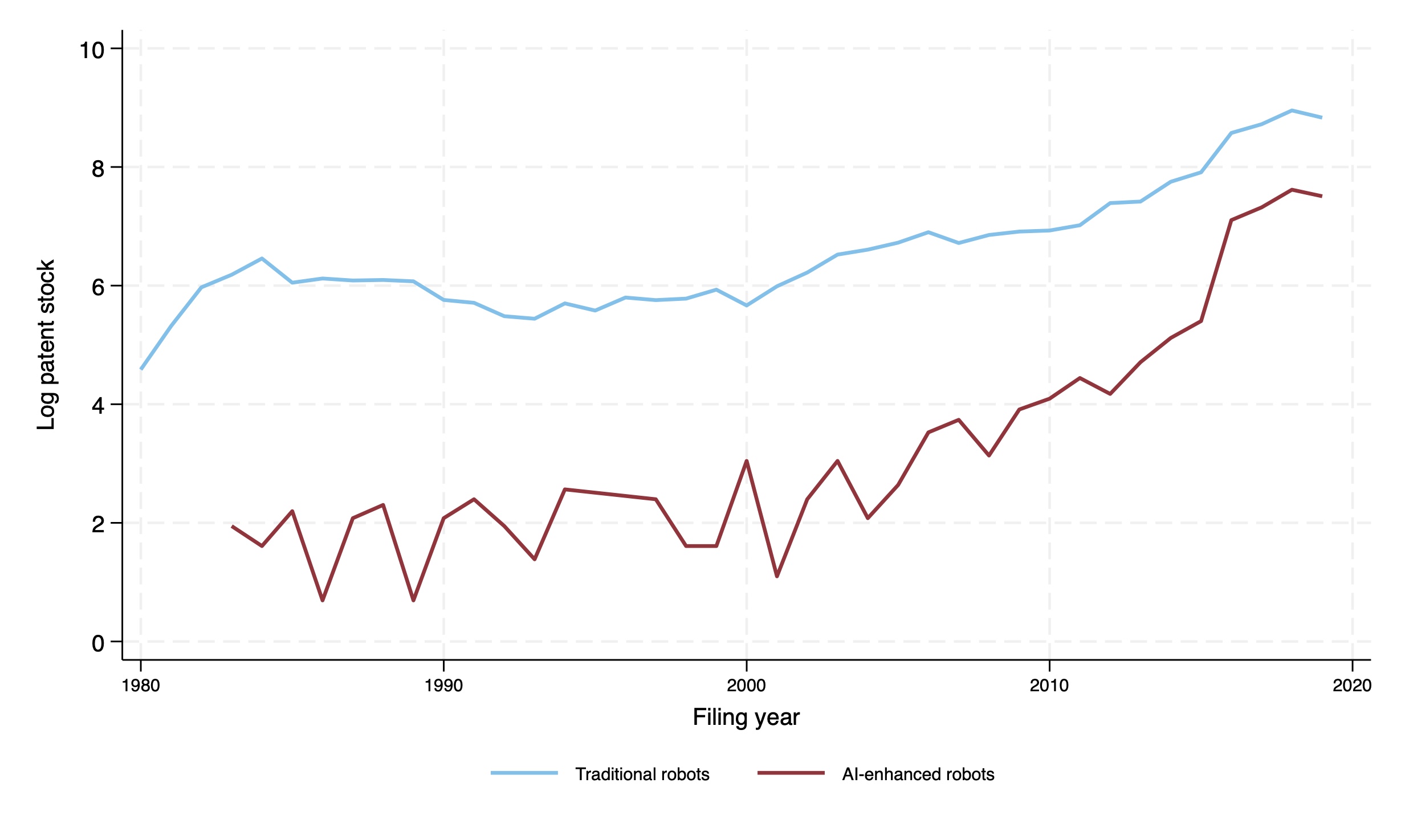}
	\floatfoot{\scriptsize
		\begin{minipage}{0.7\linewidth}
			\emph{Notes:} The figure reports the logarithm of annual patent stock in robotics, distinguishing between traditional robots (intelligent = 0) and AI-enhanced robots (intelligent = 1). 
			Patent stock is constructed from patent families and aggregated by filing year to proxy distinct inventive outputs. 
			The AI-enhanced category includes robotic inventions embedding AI-related functionalities as identified through CPC classifications, targeted keyword searches, and document-level text-mining procedures. 
		\end{minipage}
	}
\end{figure}

Turning to robotics, Figure~\ref{fig:robotpatentfamilybyyear} documents a marked increase in patent applications related to robotic technologies over the period 2015–2018, with the number of filings rising from fewer than 2,000 to nearly 6,000. While patenting activity related to traditional robots follows a relatively stable, gradual growth path over time, applications associated with AI-enhanced robots display a markedly different trajectory, characterised by modest, volatile activity in earlier years and a much steeper increase from around 2010 onwards. As a consequence, the gap between traditional and AI-enhanced robot patent stocks narrows substantially in the most recent period (Figure~\ref{fig:stockrobotslog}).

\begin{figure}[h!]
	\centering
	\caption{Shares of AI Patent Families by Filing Authority}
	\label{fig:aiauthshare}
	\includegraphics[width=0.85\linewidth]{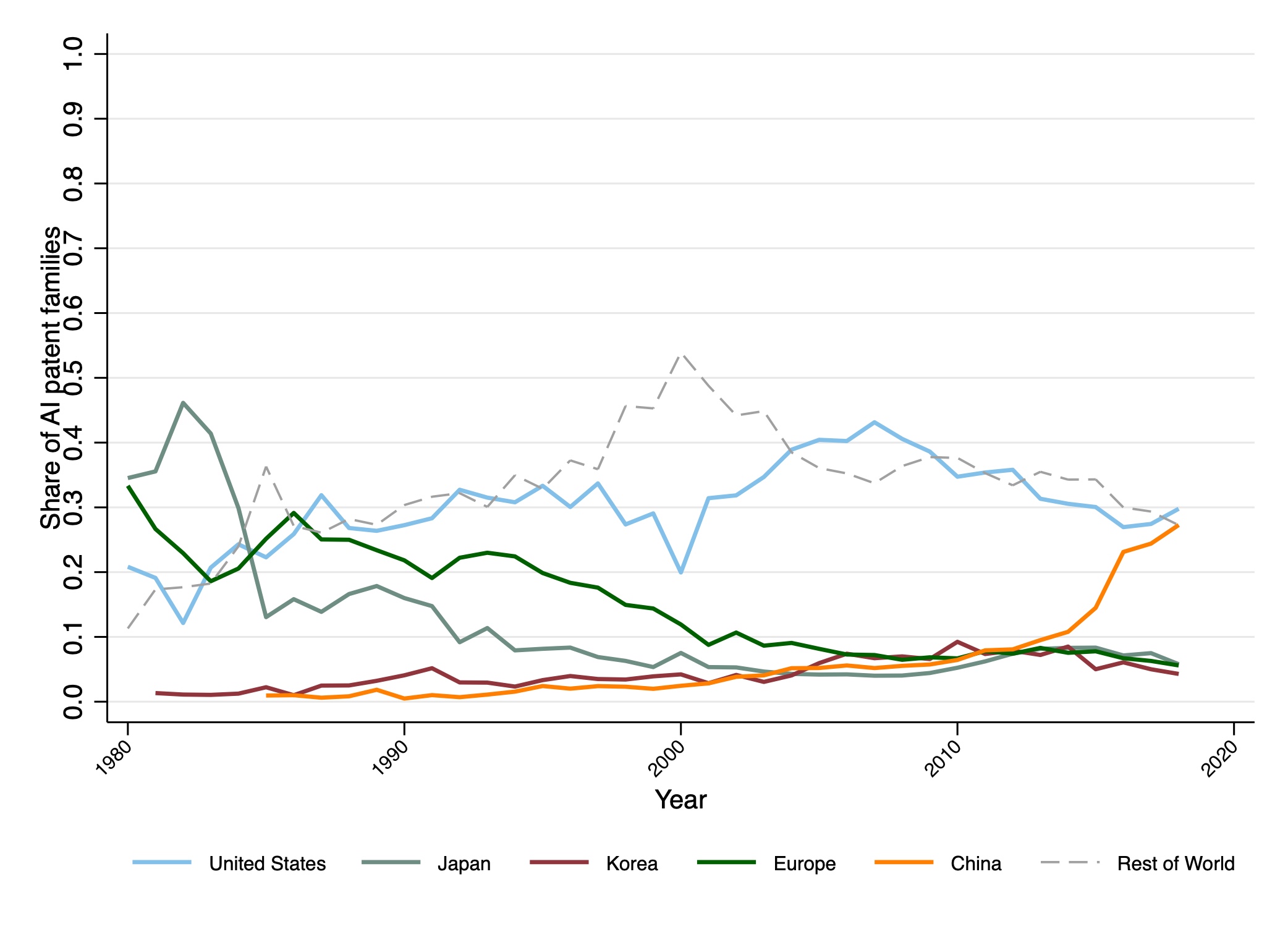}
	\floatfoot{\scriptsize
		\begin{minipage}{0.75\linewidth}
			\emph{Notes:} The figure reports, for each year and filing authority, the share of global AI patent families filed at that authority over the sum of  AI patent families across all authorities worldwide in the same year.
			Patent families are aggregated by filing year. 
			The authorities shown separately are the national patent offices of the United States (USPTO), Japan (JPO), Korea (KIPO), and China (CNIPA), as well as Europe (as defined in the dataset). 
			The Rest of the World aggregates all remaining authorities, including international applications filed under the Patent Cooperation Treaty (WIPO), regional offices (European Patent Office, Eurasian Patent Organization, African Regional Intellectual Property Organization, African Intellectual Property Organization, and European Union Intellectual Property Office), and macro-regional groupings in the dataset (Antarctica and Oceania; Americas and Caribbean; Middle East and Asia; Africa; and GC). 
			Shares sum to one within each year.
		\end{minipage}
	}
\end{figure}

Figure~\ref{fig:aiauthshare} reports the evolution of filing-authority shares in global AI patenting. In the early years, the distribution of AI patenting is relatively dispersed across major authorities, with no single economy accounting for a dominant fraction of global activity. From the early 2000s onward, the United States increases its share of worldwide AI filings. Starting in the mid-2010s, China’s share rises sharply. By contrast, the shares of Japan, Europe, and Korea remain comparatively stable or decline gradually. By the end of the sample, AI patenting is increasingly concentrated in the United States and China.

\begin{figure}[h!]
	\centering
	\caption{Shares of Robot Patent Families by Filing Authority}
	\label{fig:robotauthshare}
	\includegraphics[width=0.85\linewidth]{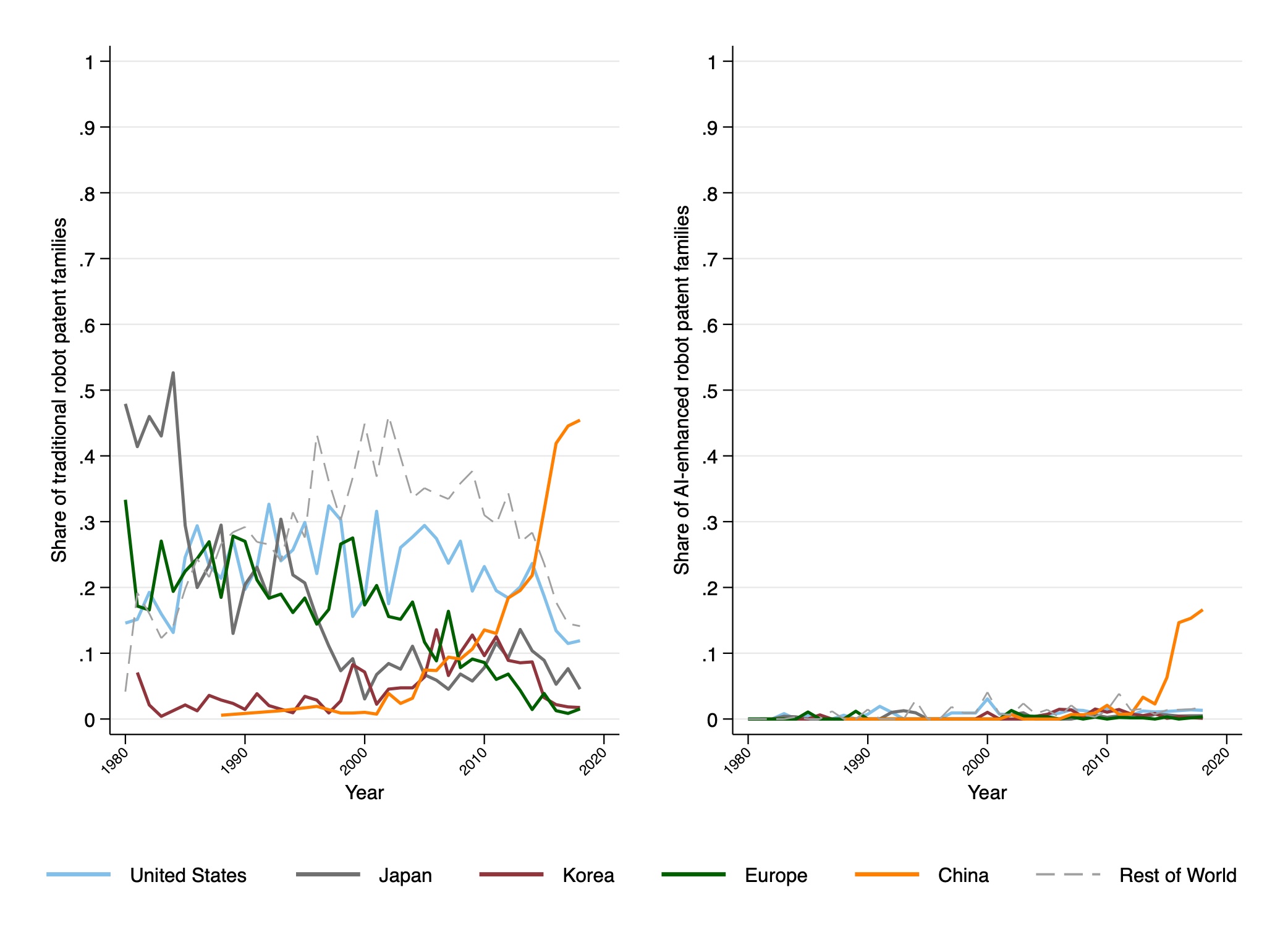}
	\floatfoot{\scriptsize
		\begin{minipage}{0.75\linewidth}
			\emph{Notes:} The figure reports, for each year and filing authority, the share of global robot patent families filed at that authority over the sum of robot patent families across all authorities worldwide in the same year. 
			Robot patent families include both traditional robots (intelligent = 0) and AI-enhanced robots (intelligent = 1) and are aggregated by filing year. 
			The authorities shown separately are national patent offices (United States/USPTO, Japan/JPO, Korea/KIPO, China/CNIPA) and Europe (as grouped in the dataset). 
			The Rest of the World aggregates all remaining authorities, including international applications filed under the Patent Cooperation Treaty (WIPO), regional offices (European Patent Office, Eurasian Patent Organization, African Regional Intellectual Property Organization, African Intellectual Property Organization, and European Union Intellectual Property Office), and macro-regional groupings in the dataset (Antarctica and Oceania; Americas and Caribbean; Middle East and Asia; Africa; and GC). 
			Shares sum to one within each year.
		\end{minipage}
	}
\end{figure}

By contrast, when considering robot-related patents in share terms, the geographical distribution of innovation evolves substantially over time (Figure~\ref{fig:robotauthshare}). In traditional robots, global patenting is initially highly concentrated in Japan, which accounts for the largest share in the 1980s. This concentration declines progressively, while the United States and Europe maintain relatively stable intermediate shares and Korea remains quantitatively smaller. China does not lead in the early period; rather, its share begins to rise markedly from the late 2000s onward and eventually exceeds that of all other authorities in the 2010s, consistent with a gradual shift in the global distribution of innovative activity.
The pattern is even more pronounced for AI-enhanced robots. Patenting activity in this technological domain remains very small across authorities until the early 2010s. Subsequently, China’s share increases sharply, while the shares of the United States, Japan, Europe, and Korea exhibit comparatively limited variation. The resulting distribution reflects a substantial rise in the geographical concentration of recent robot-related innovation.

\begin{figure}[h!]
	\centering
	\caption{Shares of AI Patent Families by Applicants’ Country}
	\label{fig:aiapplicantsshare}
	\includegraphics[width=0.85\linewidth]{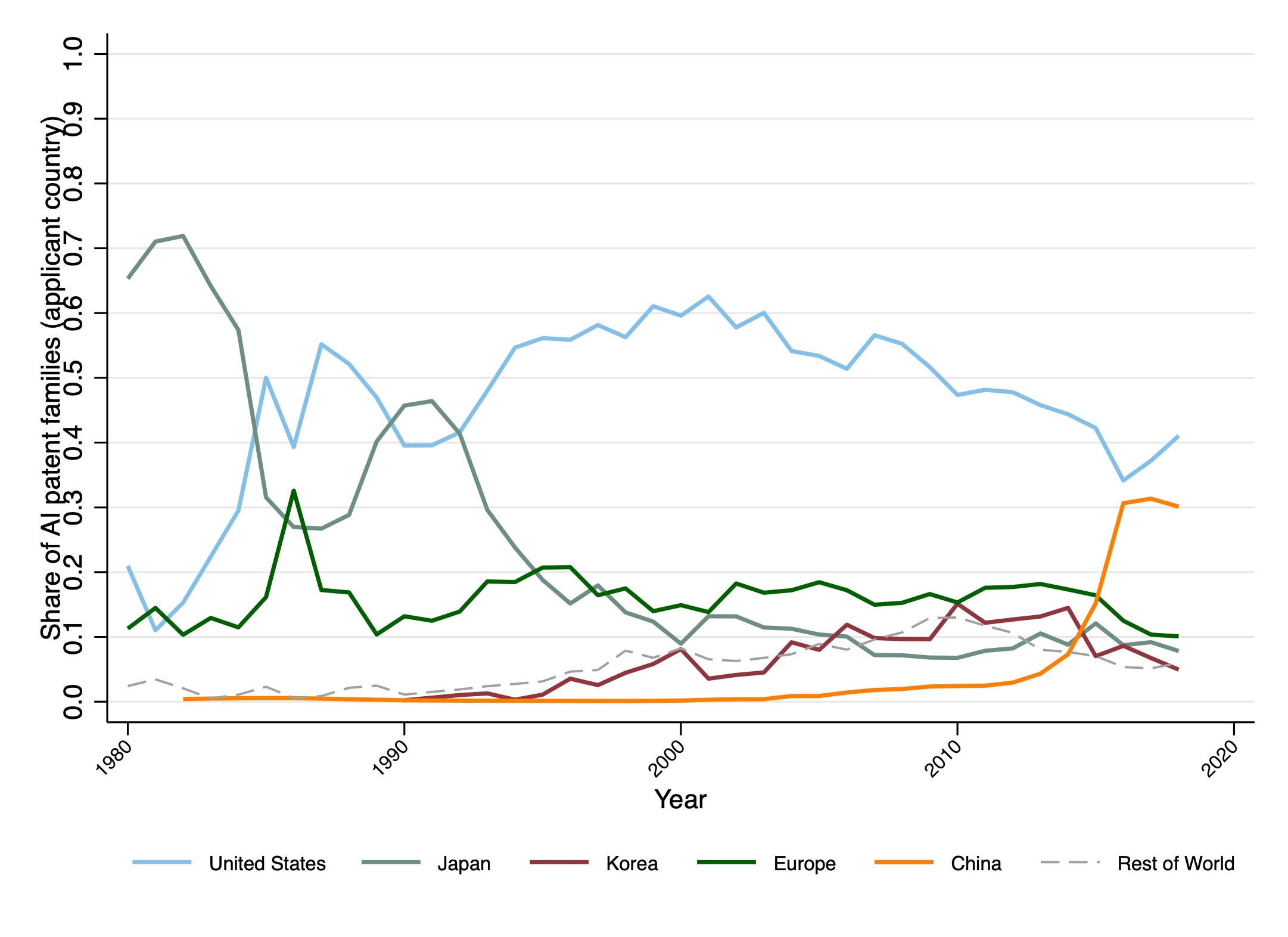}
	\floatfoot{\scriptsize
		\begin{minipage}{0.75\linewidth}
			\emph{Notes:} The figure reports applicants’ country shares of global AI patenting over the sum of AI patent families across all applicant countries worldwide in the same year. 
	       Patent families are aggregated by filing year and allocated according to the country of the applicant. 
			Shares sum to one within each year.
		\end{minipage}
	}
\end{figure}

A different pattern emerges when AI patents are allocated by applicants’ country rather than filing authority (Figure~\ref{fig:aiapplicantsshare}). In the early years, the distribution of AI patenting is relatively concentrated, with Japan accounting for a sizeable share. Over time, however, the United States becomes the dominant contributor, maintaining the largest share through the late 1990s and 2000s. Beginning in the mid-2010s, China’s share rises sharply, narrowing the gap with the United States.In contrast, the shares of Japan, Europe, and Korea exhibit comparatively limited variation over time. 

\begin{figure}[h!]
	\centering
	\caption{Applicants’ Country Shares in Robot Patent Families}
	\label{fig:noiicountriesshare}
	\includegraphics[width=0.85\linewidth]{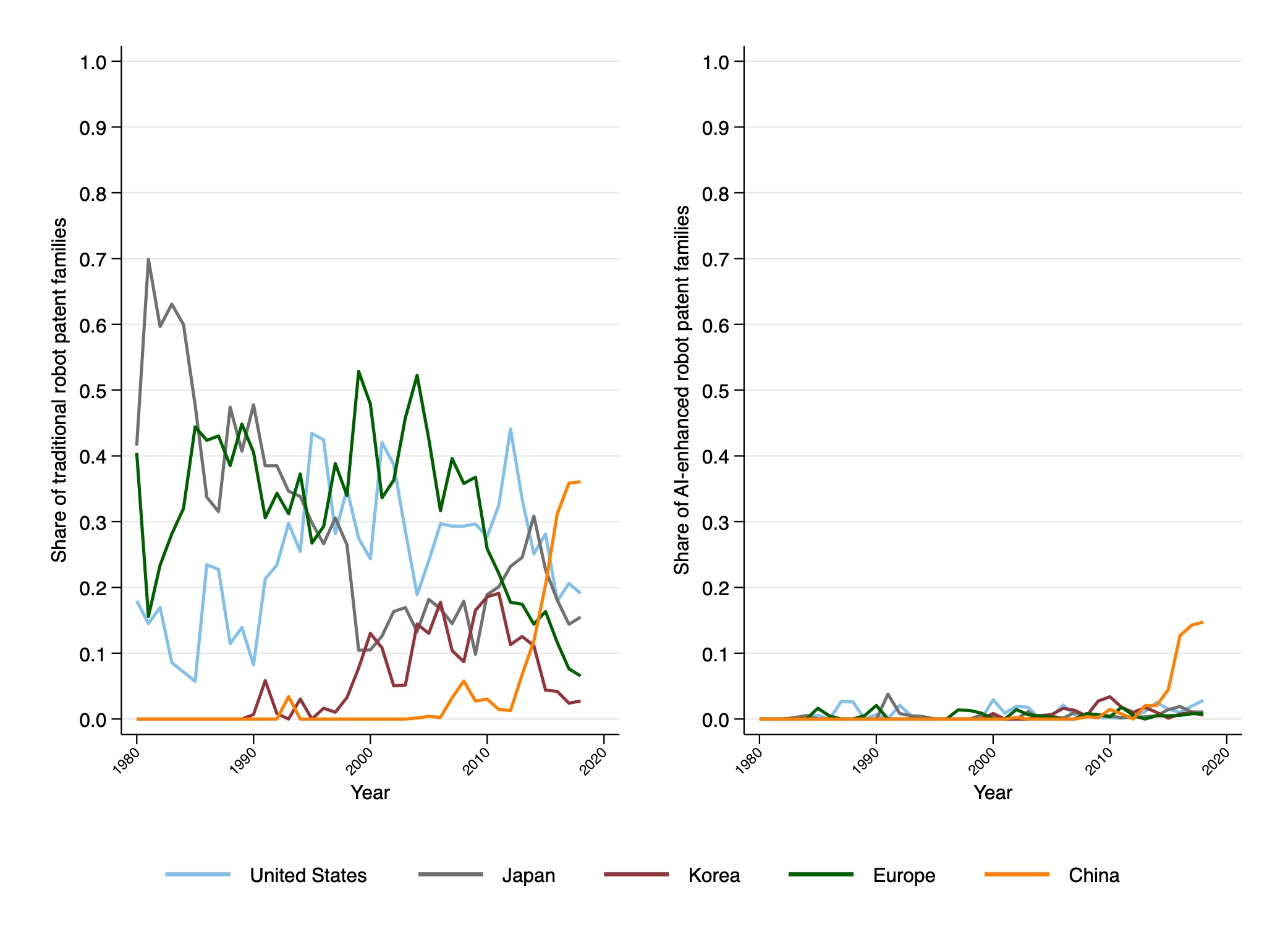}
	\floatfoot{\scriptsize
		\begin{minipage}{0.75\linewidth}
			\emph{Notes:} The figure reports applicants’ country shares in global robot patenting, distinguishing between traditional robots (NOI) and AI-enhanced robots (I). 
			For country $c$ in year $t$, shares are computed as 
			$\text{Share}_{c,t}^{k}=\frac{\text{RobotFamilies}_{c,t}^{k}}{\sum_{c'} \left(\text{RobotFamilies}_{c',t}^{NOI}+\text{RobotFamilies}_{c',t}^{I}\right)}$, 
			where $k \in \{NOI, I\}$ denotes the technological category. 
			The denominator corresponds to the total number of robot patent families worldwide in year $t$, including both traditional and AI-enhanced technologies. 
			Patent families are aggregated by filing year and assigned according to the country of the applicant. 
			Shares sum to one across countries and technologies within each year.
		\end{minipage}
	}
\end{figure}

Figure~\ref{fig:noiicountriesshare} reports applicants’ country shares in total robot patenting, distinguishing between traditional and AI-enhanced technologies. In traditional robots, Japan and Europe account for large shares in the early period, while the United States maintains a stable intermediate position throughout the sample.Over time, the decline in Japan’s share is offset by a pronounced expansion in China’s share, which accounts for most of the observed shift in the geographical distribution of traditional robot patenting.
The AI-enhanced component represents a negligible fraction of total robot patenting until the early 2010s. From the mid-2010s onward, China’s share increases sharply, whereas the U.S. share expands more modestly and does not rise proportionally.
Overall, the applicant-based perspective indicates a gradual shift in the composition of global robot innovation. While traditional robots becomes more evenly distributed over time, recent expansion is increasingly associated with China’s growing relative weight, particularly within AI-enhanced technologies.

\section{Empirical strategy}\label{sec:methods}

The empirical analysis is structured into four steps. First, statistical evidence is provided to demonstrate the non-stationarity of the time series for patents relating to AI, AI-enhanced robots and traditional robots, broken down by CPC sections. Second, if non-stationarity is present, we measure the degree of integration of the time series. Third, to model the CPC time series, we adopt a pure time-series approach. In this approach, the present values of an integrated time series depend on past values (the auto-regressive component) and a weighted average of a white noise process (the moving-average component). The first three stages follow a descriptive time series approach. The fourth stage of the empirical analysis is based on cointegration analysis. Two or more variables are defined as cointegrated if they share a common trend. This means the variables share the same long-run dynamics and jointly coevolve.

\subsection{Stationarity Analysis, Structural Breaks and the Short-Run Dynamics of Time Series}

When the probability distribution of a stochastic process remains unchanged over time, the process is strictly stationary \citep{Verbeek2017}. As stated by \citet{Pesaran2015}, this means that the realisations of a strictly stationary stochastic variable are similar over time. This implies that the probability distribution of $Y_1$ is the same as that of $Y_2$, $Y_3$ and, in general, of $Y_t$. 

The graphs described in paragraph~\ref{para:aioverview} suggest not only that the time series are non-stationary but also that there is a sudden change in the time dynamics in the neighbourhood of one or more specific years.
To test for stationarity, we use two tests: the Augmented Dickey-Fuller test and the KPSS test. 

Once we have tested for time series stationarity, we control for the presence of structural breaks. We define a structural break as an abrupt change in the profile of a time series. For this purpose, we run the Bai and Perron multiple structural test. 

However, neither short nor medium-term dynamics are revealed by non-stationarity and the presence of structural breaks.
To describe the short-run dynamics of each time series, we estimate an autoregressive ($AR$) moving average ($MA$) model for each time series. An $ARMA(p,q)$ model consists of an $AR$ component of order $p$ and an $MA$ part of order $q$ and is given by the following equation:

\begin{equation}\label{eq:arma}
y_t = \alpha_1 Y_{t-1}  + \dots + \alpha_p Y_{t-p} + \beta_1 e_{t-1} + \dots + \beta_q e_{t-q} + e_t
\end{equation}

If a time series $y_t$ is modelled using an ARMA model, this means that the time series is made up of two distinct components that coexist. The AR (auto-regressive) component indicates that the value of the time series at time $t$ depends on its past values. In contrast, the MA component implies that the value at time $t$ depends not only on the current value of the random variable but also on its past values. The key difference between these two components is that, in an AR process, a shock affects the variable's values for infinite future lags, whereas, in an MA process, a shock affects the variable for a lag depending on the order of the process. 

Therefore, to focus on the dynamics of short-term time series, we proceed as follows. First, we transform each time series to achieve stationarity. This is done by differencing the series one or more times until its statistical properties remain stable over time, yielding a stationary series. Next, we estimate the ARIMA model that best fits each stationary series. In this context, the parameter $I$ indicates the number of first differences required to achieve stationarity and is computed as the difference between consecutive terms in the original series.
To find the ARIMA equation that best fits each series, we calculate Akaike's Information Criterion (AIC), Schwarz's Information Criterion (BIC) and Hannan–Quinn's Information Criterion (HQC). These criteria indicate the optimal lag order for both the AR and MA components. 

\subsection{Cointegration analysis}\label{sec:cointegration}

After establishing stationarity and characterising the short-run dynamics of our time series, we turn to their medium- to long-term behaviour. This is examined through cointegration analysis, which allows us to identify stable long-run relationships among the series.

Two variables are said to be cointegrated when they share a common trend. 
Let us assume that two variables, $X_t$ and $Y_t$, are not stationary and are integrated of order $1$, Then we run the following cointegration regression:

\begin{equation}\label{eq:coint}
Y_t = a + BX_t + e_t
\end{equation}

If $X_t$ and $Y_t$ are cointegrated, then the error term of the estimate of the cointegrating regression will be integrated of order zero, $I(0)$. The intuition is that these two non-stationary variables can give rise to a stationary linear combination. This implies that these two variables are linked by a long-term relationship driven by an error-correction mechanism, which governs the short-run dynamics of the two series. 
In summary, this means that when two variables are cointegrated, a long-run equilibrium prevails, and this equilibrium for Equation \ref{eq:coint} is given by the following relationship: $Y_t = a + Bx$.
Furthermore, one can conceive of an error-correction representation of these data, and through these error-correction dynamics, estimate a short-run relationship consistent with the cointegration relationship. 

We follow the standard procedure based on the classical Dickey-Fuller approach to test for cointegration relationships among variables. First of all, for each country, we run the time series regression in pairs on the three time series: a) artificial intelligence, b) AI-enhanced Robots, and c) traditional robots; then we test the stationarity of residuals of each regression through the Augmented Dickey-Fuller test for unit root. 
In this way, we find that the cointegrating regressions and their residuals are used to test, using the augmented Dickey-Fuller statistics, the null hypothesis of no cointegration.%
\footnote{We have to consider that, with the exception of data about the US, our time series contains a few zeros, which could bias our estimates. Therefore, the cointegrating regressions will include a set of dummy variables that take the value of 0 when the data are nil and 1 when the values are strictly positive.} 
In order to select the optimum lag order to apply in the Dickey-Fuller test, we perform the Akaike's information criterion (AIC) and the Hannan and Quinn information criterion (HQIC).

\section{Empirical Results}\label{sec:results}

\subsection{Stationarity Analysis}

Following the procedure outlined in the previous paragraph, in Table \ref{stationarity} we show the results of the stationarity analysis. 
The tests clearly indicate that the variables A, B, G, F, and Y (representing the CPC sections of the patents, in logarithms) are non-stationary and integrated of order 1, I(1). Due to the limited length of the time series, the test could not be conducted for section E.

Two distinct groups of variables can be identified. Sections F, G, and H behave as white noise and are integrated of order 1 (I(1)), whereas the dynamics of sections A, B, and Y are better captured by ARMA models, also integrated of order 1.

\subsection{Structural Breaks}

To assess whether the observed dynamics reflect discrete shifts in technological trajectories, we test for structural breaks using the Bai–Perron methodology. Table \ref{break} reports the estimated structural breaks across CPC classes for core AI, AI-enhanced robots, and traditional robots. Two patterns emerge. First, breaks for core AI and AI-enhanced robots cluster tightly around 2010–2011 across most technological sections (A, B, E, F, G, H), indicating a common shift in innovation dynamics consistent with the diffusion of modern AI techniques.	Second, traditional robots exhibit earlier breaks in several domains, particularly in mechanical engineering (1991), physics and electricity (1999), and new technologies (1988, 1997), reflecting earlier automation waves. Although a break appears around 2015, it is less uniformly aligned with the AI-related shift. The Y (new technologies) class is particularly informative: AI shows breaks in 2006 and 2010, AI-enhanced robots in 2005 and 2010, and traditional robots much earlier (1988, 1997).  Overall, the timing of structural changes suggests that AI-enhanced robots follows the AI cycle rather than the historical trajectory of traditional robots. Table~\ref{break_inst} reports selected institutional and technological milestones identified ex post in proximity to the estimated break years, providing contextual interpretation of the findings.

\subsection{Cointegration Analysis}

The cointegration analysis is conducted for the five main regions contributing patents in AI, AI-enhanced robots, and traditional robots: China, the US, Japan, South Korea, and the EU. For each region, the analysis is performed at two levels: (a) the authority where the patents are filed, and (b) the location of the applicant's head office. Before examining the country-level dynamics, we provide a comprehensive overview of worldwide patenting activity across AI, AI-enhanced robots, and traditional robots. Finally, the analysis is carried out within each CPC class to capture sector-specific patterns, at both the filing authority level (Table \ref{DF_authorities}) and applicants' country (Table \ref{DF_applicants}).

For China, the hypothesis of no cointegration is rejected both between AI and AI-enhanced robots and between AI-enhanced robots and traditional robots, indicating a long-run relationship in these cases. In contrast, the test between AI and traditional robots does not show evidence of cointegration. These results hold consistently across analyses based on both the patent authority and the applicant's location.

The results for the US show a markedly different scenario. At the authority level, AI and AI-enhanced robots are not cointegrated, nor are AI and traditional robots. In contrast, AI-enhanced robots and traditional robots exhibit cointegration, significant at the 5\% level but not at 1\%. At the applicant level, no cointegration relationship is observed among any of the three variables.

Japan presents a distinct picture. At the authority level, only AI and AI-enhanced robots are cointegrated, significant at the 5\% level but not at 1\%. In contrast, neither AI and traditional robots nor AI-enhanced and traditional robots exhibit cointegration. At the applicant level, the results mirror those of the US, with no cointegration observed among any of the three variables.

The analysis for the EU reveals a different pattern. At the authority level, AI and traditional robots exhibit cointegration, as do AI-enhanced robots and traditional robots, while AI and AI-enhanced robots do not show evidence of cointegration. When the analysis is conducted at the applicant level, the pattern resembles that observed for the US and Japan.

Finally, the analysis of South Korean data shows that at the authority level, no evidence of cointegration is found among the variables of interest. At the applicant level, AI and AI-enhanced robots are not cointegrated, while AI and traditional robots show cointegration at the 5\% significance level (but not at 1\%). AI-enhanced robots and traditional robots do not exhibit cointegration.

To provide a more detailed discussion, we conducted the cointegration analysis by CPC section (Table \ref{DF_CPP_total}) and by country (\ref{DF_CPC_countries}).%
\footnote{The analysis is performed at the authority level only, as a high incidence of missing data prevents carrying it out at the applicant level.}
As before, the analysis was carried out for all pairs among AI, AI-enhanced robots, and traditional robots.

No evidence of cointegration is found for CPC sections A, B, and Y in China. For section F, only AI and AI-enhanced robots are significantly cointegrated. In section G, AI-enhanced robots and traditional robots show significant cointegration, while AI and AI-enhanced robots are cointegrated at a 5\% significance level. Finally, for section H, both AI and AI-enhanced robots, as well as AI and traditional robots, exhibit cointegration, albeit only at the 5\% significance level.

For the US, the cointegration patterns by CPC section differ markedly from those observed in China. In section A, no evidence of cointegration is found among the three types of patents. Section B shows cointegration only between traditional robots and AI-enhanced robots. In section F, all three pairwise combinations of time series exhibit statistically significant cointegration. Finally, for sections G and Y, traditional robots and AI-enhanced robots are significantly cointegrated.

Analysis of the data for Japan reveals a distinct pattern across CPC sections. In section A, no evidence of cointegration is found. Section B shows significant cointegration at 5\% between AI-enhanced and traditional robots. In section F, both AI and AI-enhanced robots as well as AI and traditional robots exhibit cointegration. Section G presents significant cointegration only between AI-enhanced and traditional robots. For section H, AI and traditional robots are cointegrated at 5\% significance. Finally, in section Y, both the pairs AI and traditional robots, and AI-enhanced and traditional robots, show evidence of cointegration.

For the EU, sections F and G do not show evidence of cointegration among the time series. In section A, AI-enhanced and traditional robots are cointegrated. Section B exhibits cointegration between AI and AI-enhanced robots at 5\% significance. In section H, AI and AI-enhanced robots are cointegrated at 5\% significance. Finally, in section Y, both the pairs AI and AI-enhanced robots, and AI and traditional robots, present evidence of cointegration.

Another specific set of relationships emerges for South Korea. Sections F and Y do not present evidence of cointegration among the various time series. For section A, there is evidence of cointegration between AI-enhanced robots and traditional robots, which is significant at 5\%. For section G, AI-enhanced robots and traditional robots are significantly cointegrated. Finally, in section H, both the pair AI and AI-enhanced robots and the pair AI and traditional robots present evidence of cointegration.

\subsubsection{Discussion of Empirical Findings}

Building on the previous section’s results on patenting dynamics and cointegration, we now offer an interpretative reading of these findings in the context of different innovation systems, following the framework proposed by \citet{JacobidesEtAl2021}.

First, within this framework, the US system is based on market transactions and a hyper-competitive environment. In this system, the government injects a considerable amount of resources but does not interfere in the processes of creating and destroying firms and setting up joint ventures. 
This non-interventionist government approach has favoured remarkable acquisition processes leading to the creation of tech giants (Google, Amazon, Microsoft). In such a context, we would expect to observe a low level of integration of research and patenting activities due to the low level of coordination between start-ups, incumbents and the government. 
Second, the Chinese innovation system is highly state-driven. The government not only drives the dynamics of technological change by injecting conspicuous flows of resources but also intervenes directly in the dynamics of firm creation and destruction, as well as promoting vertical integration among firms through targeted incentives. Moreover, the involvement of the universities with the private sector is more incentivised.
In this way, incumbents are not swallowed up by tech giants, as in the US, but they develop partnerships with them. These policies have also facilitated the rise of tech giants in China, but the process is strictly centrally controlled. 
Thirdly, there is the European model, where incumbents, partly due to the lack of adequate availability of venture capital, manage to survive by integrating AI capabilities through collaboration with tech giants and the creation of an ad hoc ecosystem managed by the incumbent itself.  
Finally, the Japanese and South Korean systems can be broadly, and to a first approximation, viewed as part of an Asian innovation model, where interactions among firms, government, and market forces are mediated by both institutional support and competitive dynamics.
On the one hand, the cointegration analysis largely supports Jacobides’ conceptual framework. On the other hand, it extends his conclusions by incorporating evidence from two additional key countries in AI and robotics patenting: Japan and South Korea.

Based on the cointegration analysis of time series broken down by applicant, it emerges that the Chinese innovation system, as expected, exhibits a high degree of integration in research activities across AI and robotics among firms operating in these sectors. In contrast, the cointegration analysis at the firm level indicates a relatively low degree of integration among American firms regarding AI and robotics research. The European system lies somewhere in between the Chinese and American models. Compared to the US, European markets are characterised by a lower level of competitiveness, favouring incumbents' survival. Moreover, the degree of government regulatory intervention in this area is considerably less pronounced than that observed in China. Regarding Japan and South Korea, the former resembles the Chinese system, whereas the latter aligns more with the American system. Overall, at the applicant level, cointegration analysis of these diverse institutional innovation systems highlights that the Chinese innovation system strongly integrates AI patenting with AI-enhanced robots, whereas such integration is less pronounced in the European and US systems.

Overall, our analysis at the applicant level offers an interpretative perspective on the internal dynamics of innovation systems, whereas the analysis at the authority level provides complementary evidence regarding the system’s attractiveness to innovators. Compared to the applicant-level findings, the cointegration analysis at the authority level revels additional interesting patterns. Specifically, a significant degree of cointegration is observed for CPC F (Mechanical Engineering) in both the US and Japan, while CPC G (Physics) shows a notable level of cointegration in China. In South Korea, CPC H (Electricity) exhibits a remarkable degree of cointegration, and in Europe, CPC Y (New Technologies) displays statistically significant cointegration.

\section{Conclusion}~\label{sec:conclusion}

This paper develops a patent-based framework to study the joint evolution of artificial intelligence and robots over the period 1980–2019.
By constructing a novel patent-family dataset, we distinguish between core AI, traditional robots, and robots embedding AI functionalities (AI-enhanced robots). This distinction allows us to move beyond aggregate measures of automation and examine directly how AI becomes embodied in robotic technologies. Moreover, by adopting time-series techniques, we analyse the long-run dynamics of these domains and test whether they share common stochastic trends across national innovation systems.

Several findings emerge. First, the three domains exhibit distinct long-run trajectories. Core AI expands rapidly and unevenly across technological classes. Traditional robots follow a more gradual path. AI-enhanced robots show a marked acceleration from the early 2010s onward. These patterns are consistent with the interpretation of AI as a pervasive technology whose diffusion increasingly takes place through its integration into applied domains rather than through stand-alone inventive activity.

Second, the concentration of structural breaks after 2010 further indicates a change in the growth process of AI-related technologies. In particular, breaks for AI-enhanced robots align more closely with those of core AI than with the earlier dynamics of traditional robots. This supports the view that the recent wave of robot-related innovation is increasingly driven by AI capabilities rather than by incremental improvements in conventional automation.

Third, the analysis of long-run relationships among core AI, traditional robots, and AI-enhanced robots reveals substantial cross-country variation. The cointegration tests show that the presence of common long-run dynamics depends on both the technological domains considered and the national context. In some countries, core AI and AI-enhanced robots share a long-run relationship; in others, integration between different technological pairs is observed or not detected at all.
The empirical patterns vary significantly across countries. China exhibits relatively strong cointegration between core AI and AI-enhanced robots, whereas the United States shows weaker and less pervasive long-run integration across domains. Europe, Japan, and South Korea display intermediate and technologically differentiated configurations. These findings are consistent with cross-country variation in institutional arrangements, including differences in the relative weight of public-sector research and market-based innovation.

Taken together, the results indicate that the relationship between AI and robots is best understood as a process of differentiated technological co-evolution rather than as a uniform diffusion pattern. The long-run interaction between core AI, traditional robots, and AI-enhanced robots varies systematically across domains and national innovation systems. This suggests that the economic impact of AI as a general-purpose technology depends not only on its expansion as an inventive field but also on the institutional and sectoral contexts in which it becomes embodied. By providing a consistent empirical differentiation of these domains and analysing their long-run dynamics, the paper contributes to a more precise understanding of how pervasive technologies propagate through applied technological systems.

The results document common long-run dynamics but do not identify their underlying determinants. Understanding the mechanisms behind cross-country differences remains an important direction for future research, particularly with respect to the role of industrial structure, institutional configurations, and innovation strategies.
Several extensions follow naturally. The classification developed here can be used to examine the differential firm-level and labour-market effects of core AI and AI-enhanced robots, allowing a more precise assessment of productivity and employment implications. It also provides a basis for analysing patent quality and technological influence—for example, through forward citations—to evaluate whether AI-enhanced robots represent a more impactful segment of inventive activity. Finally, extending the dataset beyond 2019 will enable us to assess whether the acceleration observed in the early 2010s intensified in the post-COVID period and whether cross-country patterns of technological integration have further converged or diverged.

\clearpage
\FloatBarrier

\begin{spacing}{1}
\bibliographystyle{apalike}
\bibliography{AI_robots_rev}
\end{spacing}

\clearpage
\FloatBarrier

\section*{Figures}

\clearpage
\FloatBarrier

\section*{Tables}


\begin{table}[h!]
	\caption{AI Patent-families: China vs US}\label{inst_fram}
	\centering
	\begin{tabular}{lrr}
		\hline\hline
		\multicolumn{1}{l}{Sectors}&\multicolumn{1}{c}{US}&\multicolumn{0}{c}{China}\\
		\hline
		Company  & 49,2 \% & 24,34\% \\
		Company Govt non-profit & 0,03\% & 0,02\%\\
		Govt non-profit & 0,94\% &  1,01\% \\
		Govt non-profit Univ. & 0,03\% &  1,34\%\\
		Hospital&0,04\% &  0,13\% \\
		Individual &47,62\% &  50,48\% \\
		University &2,10\% &  22,66\%\\
		\hline
		Total &100\% &  100\%\\
		\hline\hline
		\multicolumn{3}{l}{Source: our computation on PATSTAT data} \\		
	\end{tabular}
\end{table}

\begin{table}[h!]
	\centering
	\caption{Stationarity Analysis}
	\label{stationarity}
	
	\begin{tabular}{lcc}
		\hline\hline
		\textbf{Series} & \textbf{Selected ARIMA Model} & \textbf{Process} \\
		\hline
		A & ARIMA(2,1,3) &  \\
		B & ARIMA(2,1,2) &  \\
		F & ARIMA(0,1,0) & white noise \\
		G & ARIMA(0,1,0) & white noise \\
		H & ARIMA(0,1,0) & white noise \\
		Y & ARIMA(2,1,2) &  \\
		\hline\hline
	\end{tabular}
	
	\floatfoot{\scriptsize
		\begin{minipage}{0.65\linewidth}
			\emph{Notes:} The table reports the ARIMA specification selected for each time series after testing for stationarity. 
			Stationarity is assessed using the Augmented Dickey–Fuller (ADF) and KPSS tests. 
			All series are non-stationary in levels and are transformed by first differencing ($I=1$). 
			The optimal ARIMA$(p,d,q)$ model is selected using AIC, BIC, and HQC. 
			ARIMA$(0,1,0)$ corresponds to white noise in first differences.
		\end{minipage}
	}
	
\end{table}

\begin{table}[h!]
	\centering
	\caption{Estimated Structural Breaks by CPC Section}
	\label{break}
	
	\begin{tabular}{lccc}
		\hline\hline
		\textbf{CPC Section} & \textbf{AI-Enhanced Robots} & \textbf{Traditional Robots} & \textbf{Core AI} \\
		\hline
		A (Human necessities)      & 2010        & 2005, 2011, 2015 & 2005, 2015 \\
		B (Transporting)           & 2010        & 2011, 2015       & 2011, 2015 \\
		E (Construction)           & 2010        & 2015             & 2011 \\
		F (Mechanical engineering) & 2010        & 2011             & 1991, 2015 \\
		G (Physics)                & 2010        & 2011, 2015       & 1999, 2015 \\
		H (Electricity)            & 2010        & 2002, 2011, 2015 & 1999, 2015 \\
		Y (New technologies)       & 2006, 2010  & 2005, 2010       & 1988, 1997 \\
		\hline\hline
	\end{tabular}
	
	\floatfoot{\scriptsize
		\begin{minipage}{\linewidth}
			\emph{Notes:} The table reports the estimated break dates obtained using the Bai--Perron multiple structural break test. 
			Structural breaks are defined as statistically significant changes in the profile of the time series. 
		\end{minipage}
	}
	
\end{table}

\begin{table}[h!]
	\centering
	\caption{Structural Breaks --- Institutional Determinants}
	\label{break_inst}
	
	\scriptsize
	\setlength{\tabcolsep}{5pt}
	\renewcommand{\arraystretch}{1.15}
	
	\begin{tabularx}{\textwidth}{l l X}
		\hline\hline
		\multicolumn{3}{l}{\textbf{Panel A: AI patent families}} \\[0.4em]
		\hline
		\textbf{Country} & \textbf{Break} & \textbf{Illustrative breakthroughs} \\
		\hline
		\textbf{US}    & 2000 & Humanoid robots; driverless cars \\
		& 2015 & Formation of OpenAI \\
		\textbf{China} & 2015 & State Council: ``Made in China 2025'' \\
		\textbf{Europe}& 2014 & DeepMind acquisition (Google) \\
		\textbf{Japan} & 2011 & AI systems supporting disaster response (Tohoku) \\
		& 2015 & Robot Revolution Initiative Council \\
		\hline
		\\[-0.3em]
		\multicolumn{3}{l}{\textbf{Panel B: AI-enhanced robot patent families}} \\[0.4em]
		\hline
		\textbf{Country} & \textbf{Break} & \textbf{Illustrative breakthroughs} \\
		\hline
		\textbf{US}    & 2011 & Personal robotics development phase \\
		& 2015 & Boston Dynamics (e.g., Atlas); White House report on AI and robotics \\
		\textbf{China} & 2011 & Five-Year Plan \\
		& 2015 & Robots adopted in Alibaba warehouses \\
		\textbf{Europe}& 2010 & Advances in robotics applications (cloud-based humanoids, disaster response, surgical robots) \\
		& 2015 & EU-funded Walk-Man project; DARPA Robotics Challenge \\
		\textbf{Japan} & 2011 & Honda unveils updated ASIMO \\
		& 2015 & Government ``New Robot Strategy''; intelligent robotics initiatives \\
		\textbf{South Korea} & 2005 & KAIST develops HUBO humanoid robot \\
		& 2015 & DRC-HUBO wins DARPA Robotics Challenge \\
		\hline
		\\[-0.3em]
		\multicolumn{3}{l}{\textbf{Panel C: Non AI-enhanced robot patent families}} \\[0.4em]
		\hline
		\textbf{Country} & \textbf{Break} & \textbf{Illustrative breakthroughs} \\
		\hline
		\textbf{US}    & 2011 & Boston Dynamics introduces Cheetah \\
		& 2015 & DARPA Robotics Challenge (2015) \\
		\textbf{China} & 2011 & Large-scale investment in industrial robots \\
		& 2015 & ``Made in China 2025'': industrial robots as key sector \\
		\textbf{Europe}& 2002 & Robotics applications around euro changeover \\
		& 2012 & European Robotics Public--Private Partnership (PPP) \\
		\textbf{Japan} & 1984 & Wabot-2 humanoid robot \\
		& 2011 & Great East Japan Earthquake \\
		& 2015 & Government ``New Robot Strategy'' \\
		\textbf{South Korea} & 2004 & Early HUBO developments \\
		& 2008 & Korean Robot Act \\
		\hline\hline
	\end{tabularx}
	
	\floatfoot{\scriptsize
		\begin{minipage}{\textwidth}
			\emph{Notes:} Structural break dates are identified using the Bai--Perron multiple break procedure. 
		  Reported breakthroughs are milestones selected ex post to provide contextual interpretation around the estimated break years. 
		\end{minipage}
	}
\end{table}

\begin{table}[h!]
	\centering
	\caption{Cointegration Analysis Across Filing Authorities}
	\label{DF_authorities}
	
	\small
	\setlength{\tabcolsep}{5pt}
	
	\begin{tabular}{lcccccc}
		\toprule
		& China & US & EU & Japan & S.\ Korea & Total \\
		\midrule
		AI vs.\ Enhanced Robots          & -5.069*** & -1.580  & -2.578  & -3.599** & -0.648 & -1.244 \\
		AI vs.\ Traditional Robots       & -1.580    & -1.890  & -4.537*** & -2.201 & -2.004 & -2.059 \\
		Enhanced vs.\ Traditional Robots & -3.503*** & -3.368** & -4.340*** & -2.720* & -2.855* & -3.111** \\
		\bottomrule
	\end{tabular}
	
	\floatfoot{\scriptsize
		\begin{minipage}{\linewidth}
			\emph{Notes:} The table reports Augmented Dickey--Fuller (ADF) statistics from the Engle--Granger two-step cointegration procedure applied to the I(1) series for AI, AI-enhanced robots, and traditional robots. 
			Cointegration is tested by applying the ADF test to the residuals of each pairwise regression. 
			The null hypothesis is no cointegration (unit root in the residuals). 
			Lag length selection is based on AIC and HQIC. 
			Significance levels: *** 1\%, ** 5\%, * 10\%.
		\end{minipage}
	}
	
\end{table}

\begin{table}[htbp]
	\centering
	\caption{Cointegration Tests Across Applicants' Countries}
	\label{DF_applicants}
	
	\small
	\setlength{\tabcolsep}{5pt}
	
	\begin{tabular}{lccccc}
		\toprule
		& China & US & EU & Japan & S.\ Korea \\
		\midrule
		AI vs.\ Enhanced Robots          & -4.710*** & -0.792 & -1.965 & -1.922 & -1.584 \\
		AI vs.\ Traditional Robots       & -0.806    & -1.509 & -2.789 & -1.255 & -3.569** \\
		Enhanced vs.\ Traditional Robots & -8.199*** & -0.036 & -2.478 & -2.525 & -2.782* \\
		\bottomrule
	\end{tabular}
	
	\floatfoot{\scriptsize
		\begin{minipage}{0.95\linewidth}
			\emph{Notes:} The table reports Augmented Dickey--Fuller (ADF) statistics from the Engle--Granger two-step cointegration procedure applied to the I(1) series for AI, AI-enhanced robots, and traditional robots. 
			Cointegration is tested by applying the ADF test to the residuals of each pairwise regression. 
			The null hypothesis is no cointegration (unit root in the residuals). 
			Lag length selection is based on AIC and HQIC. 
			Significance levels: *** 1\%, ** 5\%, * 10\%.
		\end{minipage}
	}
	
\end{table}

\begin{table}[h!]
	\centering
	\caption{Cointegration Tests Across CPC Sections}
	\label{DF_CPP_total}
	
	\small
	\setlength{\tabcolsep}{5pt}
	
	\begin{tabular}{lccccccc}
		\toprule
		& A & B & E & F & G & H & Y \\
		\midrule
		AI vs.\ Enhanced Robots          & -3.187**  & -2.078  & -1.730  & -1.588  & -1.334  & -3.014**  & -2.032 \\
		AI vs.\ Traditional Robots       & -2.207    & -1.342  & -1.841  & -2.503  & -3.540** & -2.794*   & -1.631 \\
		Enhanced vs.\ Traditional Robots & -2.231    & -3.549** & -2.744* & -2.754* & -3.096** & -3.213**  & -3.956*** \\
		\bottomrule
	\end{tabular}
	
	\floatfoot{\scriptsize
		\begin{minipage}{\linewidth}
			\emph{Notes:} The table reports Augmented Dickey--Fuller (ADF) statistics from the Engle--Granger two-step cointegration procedure applied to the I(1) series for AI, AI-enhanced robots, and traditional robots. 
			Cointegration is tested by applying the ADF test to the residuals of each pairwise regression. 
			The null hypothesis is no cointegration (unit root in the residuals). 
			Lag length selection is based on AIC and HQIC. 
			Significance levels: *** 1\%, ** 5\%, * 10\%.
		\end{minipage}
	}
	
\end{table}

\begin{table}[htbp]
	\centering
	\caption{Cointegration Tests by CPC Section and Country}
	\label{DF_CPC_countries}
	
	\small
	\setlength{\tabcolsep}{4pt}
	
	\begin{tabular}{lcccccc}
		\toprule
		& A & B & F & G & H & Y \\
		\midrule
		
		\multicolumn{7}{l}{\textbf{China}} \\
		AI vs.\ Enhanced Robots          & -1.463  & -2.597  & -4.839*** & -3.079** & -4.664*** & N.A. \\
		AI vs.\ Traditional Robots       & -1.757  & -1.789  & -0.989    & -3.162** &  3.334    & N.A. \\
		Enhanced vs.\ Traditional Robots & -1.577  & -1.689  & -1.458    & -4.502*** & -1.070   & -2.555 \\
		\addlinespace[0.4em]
		
		\multicolumn{7}{l}{\textbf{United States}} \\
		AI vs.\ Enhanced Robots          &  0.453  & -2.806* & -2.981**  & -1.405  & -3.144** & -1.527 \\
		AI vs.\ Traditional Robots       & -2.477  & -2.426  & -3.658**  & -2.278  & -0.966   &  0.452 \\
		Enhanced vs.\ Traditional Robots &  0.247  & -3.059** & -3.954*** & -3.008** & -1.631  & -4.915*** \\
		\addlinespace[0.4em]
		
		\multicolumn{7}{l}{\textbf{Japan}} \\
		AI vs.\ Enhanced Robots          & -2.681* & -2.338  & -5.234*** & -1.861  & -1.860   & -2.595 \\
		AI vs.\ Traditional Robots       & -1.744  & -1.950  & -4.058*** & -1.295  & -3.003** & -5.264*** \\
		Enhanced vs.\ Traditional Robots & N.A.    & -3.640** & -3.750*** & -3.335** & N.A.    & -5.243*** \\
		\addlinespace[0.4em]
		
		\multicolumn{7}{l}{\textbf{Europe}} \\
		AI vs.\ Enhanced Robots          & -0.438  & -2.965* & -2.268    & -0.666  & -3.221** & -4.559*** \\
		AI vs.\ Traditional Robots       & -1.617  & -1.390  & -2.351    & -0.029  & -0.474   & -4.127*** \\
		Enhanced vs.\ Traditional Robots & -3.920*** & -2.962 & N.A.      & -0.031  & N.A.     & -1.612 \\
		\addlinespace[0.4em]
		
		\multicolumn{7}{l}{\textbf{South Korea}} \\
		AI vs.\ Enhanced Robots          & -1.431  & -2.604  & -2.560    &  0.826  & -1.889   & -1.946 \\
		AI vs.\ Traditional Robots       & -0.474  & -2.814* & -2.079    &  0.846  & -4.771*** & -2.489 \\
		Enhanced vs.\ Traditional Robots & -3.235** & -1.367 & N.A.      & -4.847*** & -4.217*** & -1.673 \\
		\bottomrule
	\end{tabular}
	
	\floatfoot{\scriptsize
		\begin{minipage}{\linewidth}
			\emph{Notes:} The table reports Augmented Dickey--Fuller (ADF) statistics from the Engle--Granger two-step cointegration procedure applied to the I(1) series for AI, AI-enhanced robots, and traditional robots within each country--CPC section cell. 
			Cointegration is tested by applying the ADF test to the residuals of each pairwise regression. 
			The null hypothesis is no cointegration (unit root in the residuals). 
			Lag length selection is based on AIC and HQIC. 
			N.A.\ denotes cells with insufficient observations to estimate the test. 
			Significance levels: *** 1\%, ** 5\%, * 10\%.
		\end{minipage}
	}
	
\end{table}


\clearpage
\FloatBarrier

\appendix

\pagebreak
\newpage

\section*{Appendix: Additional Tables}

\begin{table}[h!]
	\caption*{Table A1: WIPO’s and IPO’s searches incorporate keywords related to core AI concepts and methods}
	\label{tab:appendix1_keywords_ai}
	\centering
	\scriptsize 
	\setlength{\tabcolsep}{4pt} 
	\renewcommand{\arraystretch}{0.9} 
	\begin{tabular}{p{1\textwidth}}
		\hline\hline
		ARTIFIC+ OR COMPUTATION+ 1W INTELLIGEN+, NEURAL 1W NETWORK+, BAYES+ 1W NETWORK+, CHATBOT?, DATA 1W MINING+, DEEP 1W LEARNING+, GENETIC 1W ALGORITHM?, MACHINE 1W LEARNING+, NATURAL 1D LANGUAGE 1W (GENERATION OR PROCESSING), REINFORCEMENT 1W LEARNING, SUPERVISED 1W (LEARNING+ OR TRAINING), SEMI-SUPERVISED 1W (LEARNING+ OR TRAINING), CONNECTIONIS\#, EXPERT 1W SYSTEM?, FUZZY 1W LOGIC?, TRANSFER-LEARNING, LEARNING 3W ALGORITHM?, SUPPORT VECTOR MACHINE?, RANDOM FOREST?, DECISION TREE?, GRADIENT TREE BOOSTING, XGBOOST, ADABOOST, RANKBOOST, LOGISTIC REGRESSION, STOCHASTIC GRADIENT DESCENT, MULTILAYER PERCEPTRON?, LATENT SEMANTIC ANALYSIS, LATENT DIRICHLET ALLOCATION, MULTI-AGENT SYSTEM?, HIDDEN MARKOV MODEL?) and general computational/mathematical concepts frequently used in AI technologies (CLUSTERING, COMPUT+ CREATIVITY, DESCRIPTIVE MODEL?, INDUCTIVE REASONING, OVERFITTING, PREDICTIVE 1W (ANALYTICS OR MODEL?), TARGET 1W FUNCTION?, (TEST OR TRAINING OR VALIDATION) 1D DATA 1D SET?, BACKPROPAGATION?, SELF-LEARNING, OBJECTIVE FUNCTION?, FEATURE? SELECTION, EMBEDDING?, ACTIVE LEARNING, REGRESSION MODEL?, (STOCHASTIC OR PROBABILIST+) 2D (APPROACH+ OR TECHNIQUE? OR METHOD? OR ALGORITHM?), RECOMMEND+ SYSTEM?, (TEXT OR SPEECH OR HAND\_WRITING OR FACIAL OR FACE? OR CHARACTER?) 1W (ANALYSIS OR ANALYTIC? OR RECOGNITION?)\\
		\hline
IPO's CPC list similarly integrates keywords from patent abstracts\\
ant-colony factorization, machin$*$ high-dimensional$*$, feature$*$, particle-swarm$*$, bee-colony factorisation, input$*$, pattern-recogni$*$, fire-fly, feature engineer$*$, k-means, policy-gradient method, adversar$*$ network$*$, feature extract$*$, kernel learn$*$, q-learn$*$, artificial$*$-intelligen$*$, feature select$*$, latent-variable$*$, random-forest$*$, association rule, fuzzy-c link$*$, predict$*$, recommender system$*$, auto-encod$*$, fuzzy environment$*$, machine intelligen$*$, reinforc$*$ learn$*$, autonom$*$ comput$*$, fuzzy logic$*$, machine learn$*$, sentiment analy$*$, back-propagat$*$, fuzzy number$*$, map-reduce, sparse represent$*$, back-propogat$*$, fuzzy set$*$, memetic algorithm$*$, sparse$*$-code$*$, cognitiv$*$ comput$*$, fuzzy system$*$, multi$*$ label$*$ classif$*$, spectral cluster$*$, collaborat$*$ filter$*$, gaussian mixture model, multi$*$-objective$*$, algorithm$*$, stochastic$*$-gradient$*$, deep-belief network$*$, gaussian process, multi$*$-objective$*$ optim$*$, supervis learn$*$, deep-learn$*$, genetic program$*$, natural-gradient, support-vector machine$*$, differential$*$-evol$*$, genetic$*$ algorithm, neural-turing, swarm behav$*$, dimensional$*$-reduc$*$, high-dimensional$*$ data, neural-network, swarm intell$*$, ensemble-learn$*$, high-dimensional$*$ model$*$, neuro-morph comput$*$, transfer-learn$*$, evolution$*$ algorithm$*$, high-dimensional$*$ system$*$, object-recogni$*$, vector-machine$*$)\\
		\hline\hline
\end{tabular}
\end{table}

\begin{table}[h!]
	\caption*{Table A2: CPC Search Terms for AI}
	\label{tab:cappendix2_cpc_ai}
	\centering
	\scriptsize 
	\setlength{\tabcolsep}{4pt} 
	\renewcommand{\arraystretch}{0.9} 
	\begin{tabular}{p{0.03\textwidth}p{0.95\textwidth}}
		\toprule
		No. & Search Terms \\
		\midrule
		CPC & A61B 5/7264, A61B 5/7267, A63F 13/67, A63F2300/00, 
		B23K 31/006, B25J 9/161, B29C 66/965, B29C2945/76979, 
		B60G2600/1876, B60G2600/1878, B60G2600/1879, B60W 30/06, 
		B60W 30/10, B60W 30/12, B60W 30/14, B62D 15/0285, 
		B64C2201/00, B64C2201/02, B64C2201/021, B64C2201/024, 
		B64C2201/025, B64C2201/027, B64C2201/028, B64C2201/04, 
		B64C2201/042, B64C2201/044, B64C2201/046, B64C2201/048, 
		B64C2201/06, B64C2201/063, B64C2201/066, B64C2201/08, 
		B64C2201/082, B64C2201/084, B64C2201/086, B64C2201/088, 
		B64C2201/10, B64C2201/101, B64C2201/102, B64C2201/104, 
		B64C2201/105, B64C2201/107, B64C2201/108, B64C2201/12, 
		B64C2201/121, B64C2201/122, B64C2201/123, B64C2201/125, 
		B64C2201/126, B64C2201/127, B64C2201/128, B64C2201/14, 
		B64C2201/141, B64C2201/143, B64C2201/145, B64C2201/146, 
		B64C2201/148, B64C2201/16, B64C2201/162, B64C2201/165, 
		B64C2201/167, B64C2201/18, B64C2201/182, B64C2201/185, 
		B64C2201/187, B64C2201/20, B64C2201/201, B64C2201/203, 
		B64C2201/205, B64C2201/206, B64C2201/208, B64C2201/22, 
		B64G2001/247, E21B2041/0028, F02D 41/1405, F03D 7/046, 
		F05B2270/707, F05B2270/709, F05D2270/707, F05D2270/709, 
		F16H2061/0081, F16H2061/0084, G01N 29/4481, G01N 33/0034, 
		G01N2201/1296, G01R 31/2846, G01R 31/3651, G01S 7/417, 
		G05B 13/0265, G05B 13/027, G05B 13/0275, G05B 13/028, 
		G05B 13/0285, G05B 13/029, G05B 13/0295, G05B2219/33002, 
		G05D 1/00, G05D 1/0088, G06F 11/1476, G06F 11/2257, 
		G06F 11/2263, G06F 15/18, G06F 17/14, G06F 17/153, 
		G06F 17/16, G06F 17/2282, G06F 17/27, G06F 17/28, 
		G06F 17/30, G06F 17/50, G06F 19/24, G06F 19/707, 
		G06F2207/4824, G06K 7/1482, G06K 9/00, G06K 9/00791, 
		G06N 3/00, G06N 3/004, G06N 3/02, G06N 3/12, 
		G06N 5/00, G06N 5/003, G06N 5/006, G06N 5/02, 
		G06N 5/022, G06N 5/025, G06N 5/027, G06N 5/04, 
		G06N 5/041, G06N 5/042, G06N 5/043, G06N 5/045, 
		G06N 5/046, G06N 5/047, G06N 5/048, G06N 7/00, 
		G06N 7/005, G06N 7/02, G06N 7/023, G06N 7/026, 
		G06N 7/04, G06N 7/043, G06N 7/046, G06N 7/06, 
		G06N 7/08, G06N 20/00, G06N 20/10, G06N 20/20, 
		G06N 99/005, G06Q 30/02, G06T 1/20, G06T 3/4046, 
		G06T 7/00, G06T 9/002, G06T2207/20081, G06T2207/20084, 
		G06T2207/30236, G06T2207/30248, G08B 29/186, G10H2250/005, 
		G10H2250/151, G10H2250/311, G10K2210/3024, G10K2210/3038, 
		G10L 13/00, G10L 15/00, G10L 17/00, G10L 25/00, 
		G10L 25/30, G10L 99/00, G11B 20/10518, G16H 50/20, 
		H01J2237/30427, H02P 21/0014, H02P 23/0018, H03H2017/0208, 
		H03H2222/04, H04L 25/0254, H04L 25/03165, H04L 41/16, 
		H04L 45/08, H04L2012/5686, H04L2025/03464, H04N 21/4662, 
		H04N 21/4663, H04N 21/4665, H04N 21/4666, H04Q2213/054, 
		H04Q2213/13343, H04R 25/507, Y10S 128/924, Y10S 128/925, 
		Y10S 706/00, Y10S 706/90, Y10S 706/902, Y10S 706/903, 
		Y10S 706/904, Y10S 706/905, Y10S 706/906, Y10S 706/907, 
		Y10S 706/908, Y10S 706/909, Y10S 706/91, Y10S 706/911, 
		Y10S 706/912, Y10S 706/913, Y10S 706/914, Y10S 706/915, 
		Y10S 706/916, Y10S 706/917, Y10S 706/918, Y10S 706/919, 
		Y10S 706/92, Y10S 706/921, Y10S 706/922, Y10S 706/923, 
		Y10S 706/924, Y10S 706/925, Y10S 706/926, Y10S 706/927, 
		Y10S 706/928, Y10S 706/929, Y10S 706/93, Y10S 706/931, 
		Y10S 706/932, Y10S 706/933, Y10S 706/934 \\
		\bottomrule
	\end{tabular}
\end{table}

\begin{table}[h!]
	\caption*{Table A3: Search strategy for robots (CPC, keyword-based, and text-mining approach)}
	\label{tab:appendix3_robot_search_terms}
	\centering
	\scriptsize
	\setlength{\tabcolsep}{6pt}
	\renewcommand{\arraystretch}{1.15}
	
	\begin{tabular}{p{0.22\textwidth}p{0.74\textwidth}}
		\toprule
		\textbf{Category} & \textbf{Search terms} \\
		\midrule
		
		\textbf{CPC} &
		B25J \\
		\midrule
		
		\textbf{Robot keywords} &
		\textit{Industrial robot}, \textit{service robot}, \textit{social robot}, \textit{robot system} \\
		\midrule
		
		\textbf{Co-occurrence filter} &
		\textit{Cognitive system}, \textit{control theory} \\
		\midrule
		
		\textbf{Text-mining} &
		\textbf{Learning \& intelligence:} artificial intelligence; intelligent; intelligence; smart; learning; Machine Learning; reasoning; cognitive; neural network; support vector; vector machine; gradient descent; q-learning; bayesian optimisation; bayesian modelling; bayesian controllers; probabilistic model; gaussian process; random-forest; factorization machines; factorisation machines; adversial network; automated theorem proving; inductive programming; case-based reasoning; constraint satisfaction; latent variable models; causal models; genetic algorithm; genetic programming control; evolutionary algorithm; evolutionary computation; metaheuristic optimisation; stochastic optimisation; bagging; boosting; AI analysis; AI application; AI benchmark; AI software toolkit; computational creativity; computational linguistics; computational economics; negotiation algorithm; game theory; question-answering robot system; Question-answering robot system; text mining; text classification; automatic classification.
		
		\vspace{0.3em}
		\textbf{Perception \& recognition:}
		computer vision; image processing; image recognition; image retrieval; object recognition; object detection; object classification; pattern recognition; automatic recognition; face recognition; facial recognition; gesture recognition; speech recognition; speech processing; speech synthesis; synthetic speech; sound event recognition; sound source separation; sound synthesis; sound description; computational auditory scene; music information retrieval; visual memory; smart sensor; smart sensor system; visual sensor network; sensor tactile systems; synthetic sensor; Light Detection and Ranging; LIDAR sensors.  
		
		\vspace{0.3em}
		\textbf{Data \& connectivity:}
		big data; data mining; data analytics; network data; sensor network; wireless sensor network; Internet of Things; IoT; semantic web; cloud; Cloud; integration data; analytics platform; real-time sensor data; reality time. 
		
		\vspace{0.3em}
		\textbf{Control \& decision systems:}
		control theory; intelligent control; fuzzy logic; fuzzy control; fuzzy system; decision support; decision analytics; prediction; information-control system; AI-based controllers; artificial intelligence controller; AI-based control system; machine learning control systems; artificial intelligence control systems; learning-based control; perception controller; AI-based robots.  
		
		\vspace{0.3em}
		\textbf{Autonomy:}
	multi-agent systems; agent-based modelling; autonomous system; autonomous robot; Autonomous mobile robot; autonomous mobile robot systems; Autonomous multi-platform robotic system; autonomous vehicle; autonomous driving; self-driving car; self-driving vehicle; unmanned vehicle; unmanned aerial vehicle; swarm robot; particle swarm; swarm optimization; ant-colony; bee colony; smart robot; interaction robot; autonomous interactions; AI humanoid robots; AI-enabled robotics; robotic perception; expressive social robot; social interaction; expressive communication; character interaction; personality profile; synthetic sensor. 
		
		\vspace{0.3em}
		\textbf{Other AI-related technologies:}
		augmented reality; virtual reality; virtual environment; man-machine interaction; human-AI interaction; natural language understanding; natural language input; sentiment analysis; anomaly detection; component analysis; collaborative filtering; content-based filtering; tensor processing unit; AI accelerator; General-AI; Specific-AI; Vertical-AI; Narrow-AI; AI-; computational creativity.
		\\
		\bottomrule
	\end{tabular}
\end{table}

\end{document}